\documentclass[%
 reprint,
superscriptaddress,
 amsmath,amssymb,
 aps,
 longbibliography,
]{revtex4-1}

\usepackage{graphicx}
\usepackage{dcolumn}
\usepackage{bm}

\begin{document}

\preprint{APS/123-QED}

\title{Linker-mediated phase behavior of DNA-coated colloids}

\author{Janna Lowensohn} 
\affiliation{Martin A. Fisher School of Physics, Brandeis University, Waltham, MA 02453, USA} 

\author{Bernardo Oyarz\'un} 
\affiliation{Universit\'e Libre de Bruxelles, Interdisciplinary Center for Nonlinear Phenomena and Complex Systems, Campus Plaine, Code Postal 231, Blvd du Triomphe, B-1050 Brussels, Belgium} 

\author{Guillermo Narvaez Paliza} 
\affiliation{Martin A. Fisher School of Physics, Brandeis University, Waltham, MA 02453, USA} 

\author{Bortolo M. Mognetti}
\affiliation{Universit\'e Libre de Bruxelles, Interdisciplinary Center for Nonlinear Phenomena and Complex Systems, Campus Plaine, Code Postal 231, Blvd du Triomphe, B-1050 Brussels, Belgium} 

\author{W. Benjamin Rogers} 
\email{wrogers@brandeis.edu}
\affiliation{Martin A. Fisher School of Physics, Brandeis University, Waltham, MA 02453, USA} 

\date{\today}

\begin{abstract}
The possibility of prescribing local interactions between nano- and microscopic components that direct them to assemble in a predictable fashion is a central goal of nanotechnology research. In this article we advance a new paradigm in which self-assembly of  DNA-functionalized colloidal particles is programmed using linker oligonucleotides dispersed in solution. We find a phase diagram that is surprisingly rich compared to phase diagrams typical of other DNA-functionalized colloidal particles that interact by direct hybridization, including a re-entrant melting transition upon increasing linker concentration, and show that multiple linker species can be combined together to prescribe many interactions simultaneously. A new theory predicts the observed phase behavior quantitatively without any fitting parameters. Taken together, these experiments and model lay the groundwork for future research in programmable self-assembly, enabling the possibility of programming the hundreds of specific interactions needed to assemble fully-addressable, mesoscopic structures, while also expanding our fundamental understanding of the unique phase behavior possible in colloidal suspensions.

\end{abstract}

\maketitle

DNA-coated colloids are one of the most promising systems for designing complex self-assembling materials \cite{Rogers_2016_NatRevMat, Di_Michele_2013_PCCP,Jones_2015_Science}.  As in nature, the information required to specify the interactions and assembly pathways leading to a desired structure can be stored in the building blocks themselves. In the case of DNA-coated particles, this information is stored in the base sequences. In recent years, considerable progress has been made in using DNA to program the self-assembly of a variety of  crystalline materials \cite{Macfarlane_2011_Science,srinivasan2013designing,macfarlane2014importance, Wang_2015_NatComm,Wang_2017_NatComm}. However, experimental demonstrations of the addressable assembly of DNA-coated particles into fully prescribed structures have yet to be realized.

Recent theoretical work highlights the challenges of using DNA-coated particles for assembling prescribed materials, which need not be symmetric or periodic \cite{Hormoz_2011_PNAS,Zeravcic_2014_PNAS}. To produce an arbitrary, complex structure with high yield from particles with specific, yet isotropic interactions, every particle must be different and have interactions chosen to favor the desired local configuration of the target structure \cite{Zeravcic_2014_PNAS}. Furthermore, all favorable interactions must have comparable energies \cite{Hormoz_2011_PNAS}. As a result, programming the assembly of even modest structures, which might contain only dozens of particles, requires specifying hundreds of unique binding interactions, all of which must have the same affinity. 

\begin{figure}
\includegraphics[width=0.98\columnwidth]{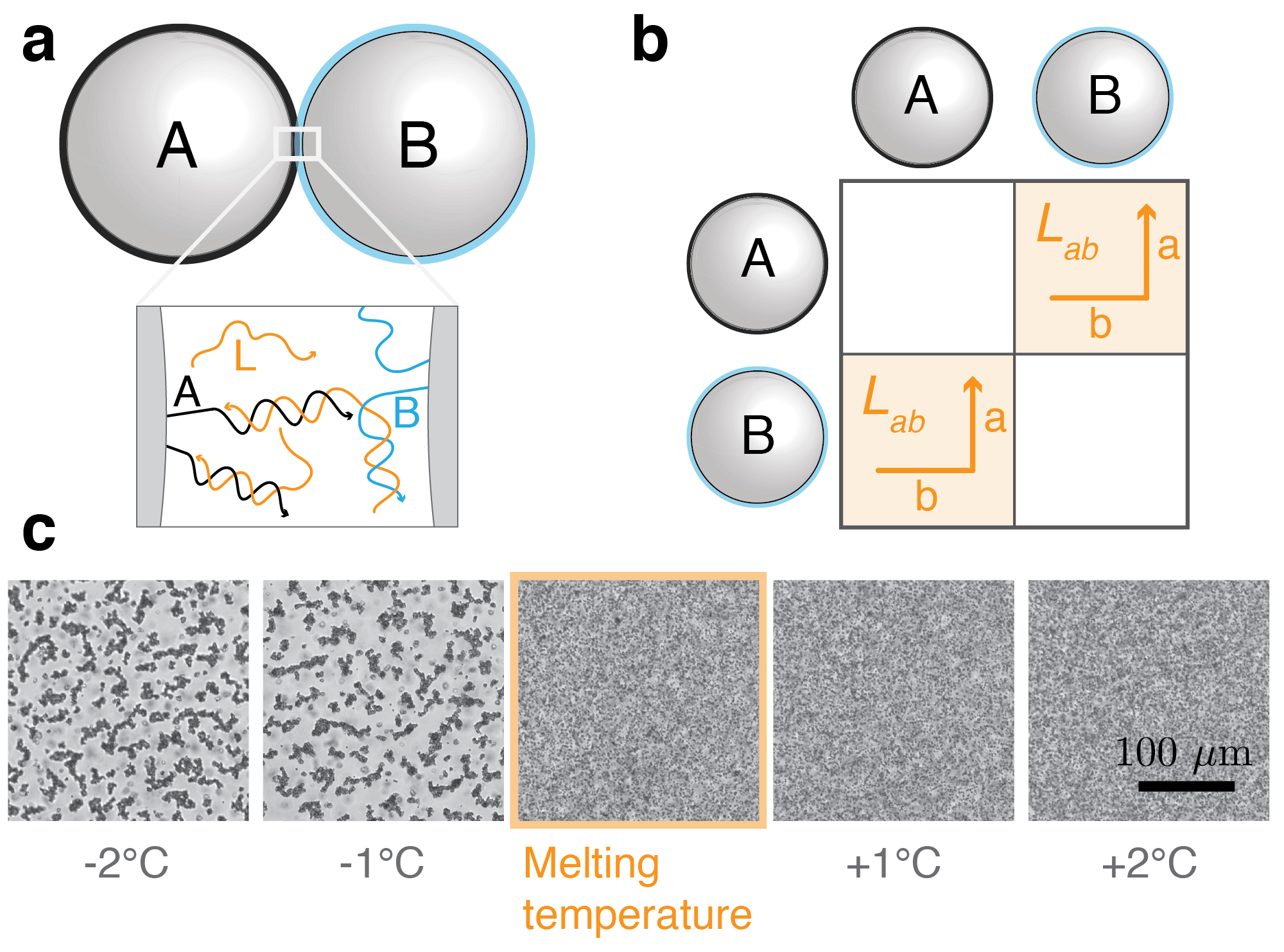}
\caption{Linker-mediated binding. (a) Our experimental system is comprised of three DNA sequences: strands A and B are grafted to 1-$\mu$m-diameter colloidal particles, and strands $L_{ab}$, which bind A to B, are dispersed in solution.  (b) These three DNA sequences produce a symmetric interaction matrix, in which the linker encodes the pair interaction between particles A and B. (c) The resulting phase behavior is temperature dependent: the system phase separates upon decreasing temperature, as shown by optical micrographs. We define the melting temperature ($T_m$) as the temperature at which 50\% of the particles are unbound. Experiments are for a 19-nucleotide linker at a concentration of 1~$\mu$M; the melting temperature is 43 degrees Centigrade. 
\label{fig:overview}}
\end{figure}

In principle, DNA can encode these hundreds of interactions through careful design of the base sequences \cite{Zeravcic_2014_PNAS,Wu_2012_PNAS}.  In practice, however, this potential is nearly impossible to realize in systems of DNA-coated particles interacting through direct binding of their grafted strands: the steep temperature dependence of the interactions \cite{Biancaniello_2005_PRL, Dreyfus_2009_PRL, Rogers_2011_PNAS}, the inherent uncertainty in predictions of the binding affinities \cite{SantaLucia_1998_PNAS}, and the inability to tune the relative interactions without resynthesizing the particles \cite{Kim_2005_JACS}, make matching hundreds of unique interactions intractable. While strategies have been explored to reduce the number of specific interactions, including adding directional binding and exploiting hierarchical pathways to assembly \cite{Halverson_2013_PRE,halverson2017communication}, these come with their own practical challenges.

An alternative approach is to design particles that interact through single-stranded DNA (ssDNA) oligomers dissolved in solution instead of through direct binding of grafted strands. Here the binding kinetics, the interaction strength, and even the interaction matrix itself could be tuned by changing the concentrations and sequences of the soluble linker strands. Furthermore, linker-mediated assembly could in principle enable the programming of large sets of specific interactions using considerably fewer unique sequences than required by systems interacting by direct hybridization. However, the practical limits of linker-based systems have not been determined and a quantitative understanding of how the linker sequences, concentrations, and surface densities of grafted molecules affect the emergent phase behavior is missing \cite{Biancaniello_2005_PRL, Xiong_2009_PRL, Xiong_2008_JACS, Tkachenko_2002_PRL}. 

In this article we combine experiments and theory to explore the phase behavior that emerges when interactions between DNA-grafted colloidal particles are encoded in soluble linker molecules. Our experiments reveal a rich phase diagram containing two previously unknown regions: (1) a re-entrant melting transition that occurs upon increasing linker concentration; and (2) a linker concentration at which coexistence between fluid and solid is stable over a wide range of temperatures. We show that the phase boundaries separating fluid and solid can be tuned by adjusting the grafting density, linker sequence, and concentration, and also demonstrate that a number of competing linker sequences can coexist in the same solution without interfering with one another, suggesting that it might be possible to encode hundreds of interactions simultaneously. Lastly, we develop a statistical-mechanical model that captures the unique phase behavior that we observe quantitatively, showing that we can predict and thus program the interactions required to direct assembly of prescribed, aperiodic structures.  

\section*{Results and discussion}
\subsection*{Linker-mediated phase behavior}
Our experimental system consists of a binary mixture of micrometer-diameter colloidal particles coated with ssDNA\cite{Kim_2005_JACS}. Each particle species bears 65-base-long, single-stranded DNA oligonucleotides which have a 54-thymine spacer and a unique 11~nucleotide~(nt) ``sticky end'', called either $A$ or $B$. The sticky ends are not directly complementary, but can be linked together by single-stranded oligonucleotides dissolved in solution, which are half complementary to $A$ and half complementary to $B$ (Fig.~\ref{fig:overview}a). We call these strands ``linkers''. 

The linker-dependent interactions can be represented in a symmetric matrix (Fig.~\ref{fig:overview}b), where each element of the matrix is encoded by a single linker sequence (e.g. $L_{ab}$ specifies the interactions between $A$ and $B$). As with binding due to direct hybridization, our linker-mediated interactions are temperature dependent: the particles aggregate when cooled and disaggregate when heated (Fig.~\ref{fig:overview}c).  We characterize the phase behavior of our system using the melting temperature $T_m$, which we define as the temperature at which half of the particles are completely unbound \cite{Dreyfus_2009_PRL}. See Supporting Information (SI) Section S1 for experimental details.

\begin{figure}
\includegraphics[width=0.98\columnwidth]{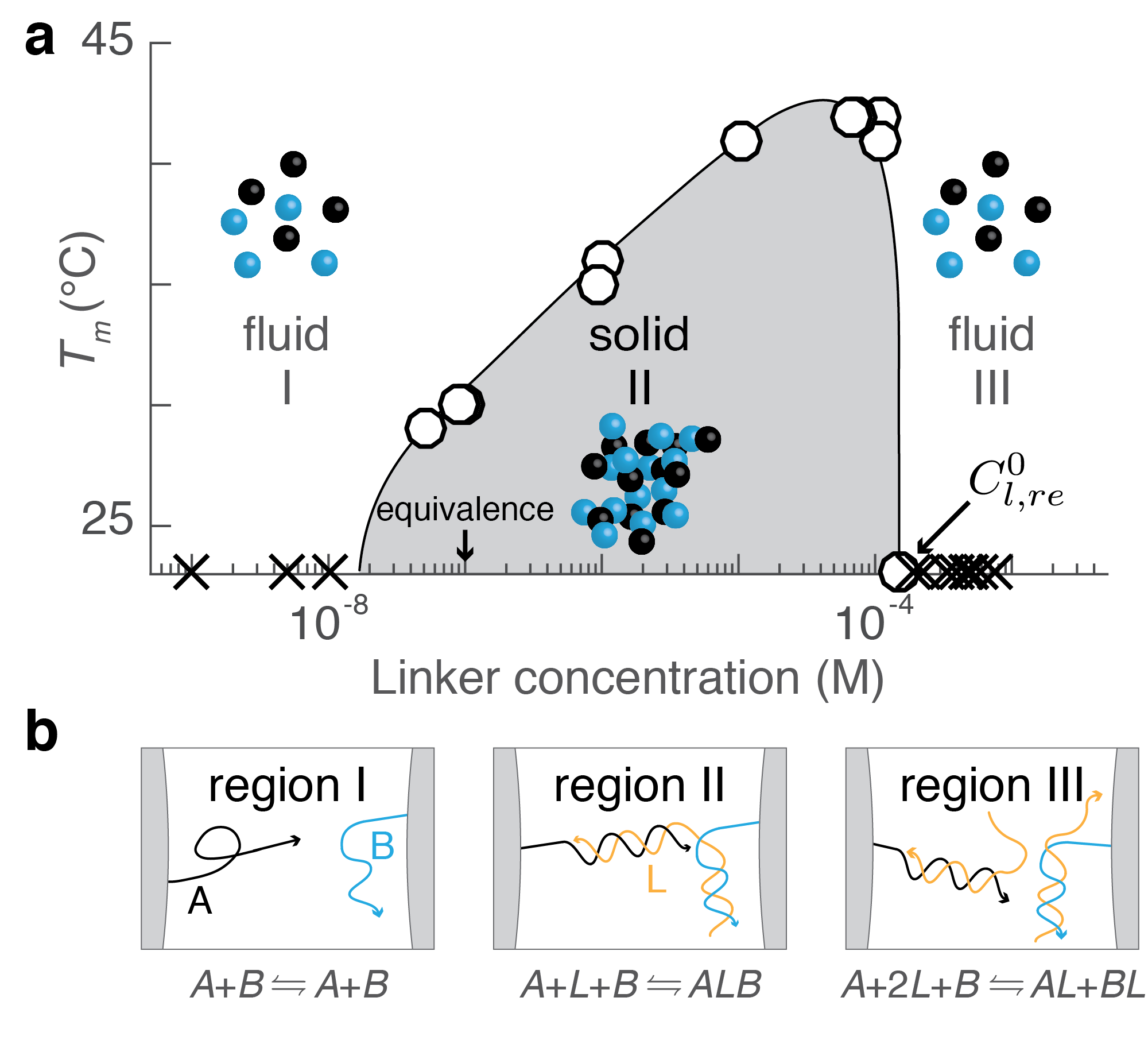}
\caption{Phase behavior of linker-mediated assembly.  (a) The temperature-linker concentration phase diagram is characterized by three distinct regimes: a fluid phase at low linker concentrations (region I), a solid-fluid transition at intermediate linker concentrations (region II), and a re-entrant fluid phase at the highest linker concentrations (region III). Circles show measurements of the melting temperatures at different linker concentrations; x's show samples that were fluid at all temperatures. We define the re-entrant concentration $C_{l,re}^0$ as the linker concentration above which the solid melts. The arrow labeled ``equivalence'' shows the linker concentration at which there is one linker molecule per grafted molecule in the system. Data are for the 17-nt linker and particles with a DNA grafting density of 2000~DNA/$\mu$m$^2$. We hypothesize that the phase behavior in (a) results from the molecular-scale reactions shown in (b).
\label{fig:phase-diagram}}
\end{figure}

The concentration-dependent phase behavior that we find is unexpectedly rich, featuring three distinct regions upon increasing linker concentration (Fig.~\ref{fig:phase-diagram}). At the lowest linker concentrations, particles do not aggregate even at room temperature (Fig.~\ref{fig:phase-diagram}a, region I). At intermediate linker concentrations, the particles aggregate at low temperatures and disaggregate when heated (Fig.~\ref{fig:phase-diagram}a, region II). Within this region of the phase diagram, the temperature at which the particles disassociate $T_m$ increases roughly monotonically with increasing linker concentration, increasing by roughly 10--15 degrees Centigrade upon increasing the linker concentration from about 100~nM to 100~$\mu$M. Above an even higher linker concentration the particles fail to aggregate completely over the entire experimental temperature range (Fig.~\ref{fig:phase-diagram}a, region III). We call the linker concentration above which particles fail to aggregate the ``re-entrant concentration'' $C_{l,re}^0$.

We can understand the observed phase behavior qualitatively by considering the molecular-scale reactions that give rise to interactions between colloidal particles (see Fig.~\ref{fig:phase-diagram}b). At the lowest linker concentrations, there are too few linkers to stabilize bridges linking particles together. At intermediate linker concentrations, linkers can form bridges between particles via the molecular-scale reaction $A + L + B\leftrightharpoons ALB$. Under these conditions, the melting temperature increases upon increasing linker concentration, since increasing the amount of linker shifts the local equilibrium toward the bridged conformation $ALB$. At the highest linker concentrations, we suspect that every grafted strand is bound to its own linker. Since the linkers cannot bind to other linkers, the particles become effectively passivated against assembly. We note that the state with two half bridges $AL$ and $BL$ has the same total number of base pairs as the bridged state $ALB$, thus we suspect that the solid phase within region II is actually stabilized by the entropy of the free linker strands dispersed in solution \cite{Rogers_2015_Science}. 

The generic phase diagram that emerges illustrates two important features of linker-mediated self-assembly: 1) there is a wide `dynamic range' of linker concentrations---spanning roughly 4 orders of magnitude---over which the melting temperature (and thus the interaction strength) can be tuned; and 2) there is a linker concentration above which colloids cannot self-assemble, irrespective of the temperature. We highlight that the re-entrant concentration is not simply the ``equivalence point'' at which there is one linker molecule for each grafted strand in the system (labeled ``equivalence'' in Fig.~\ref{fig:phase-diagram}a). Indeed, for a grafting density of 2000~DNA/$\mu$m$^2$ and a total particle volume fraction of 0.5$\%$, there are roughly 2000 linkers for each grafted DNA strand at the re-entrant concentration we find in experiment.

\begin{figure}
\includegraphics[width=0.98\columnwidth]{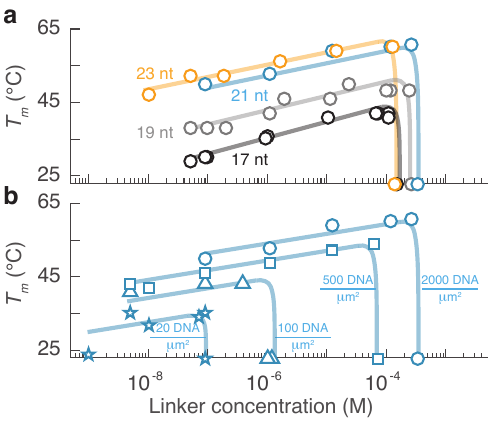}
\caption{Effects of linker affinity and grafting density on the phase behavior. (a) Experimental measurements of the melting temperature as a function of linker concentration for four linkers having different lengths $n$: 17~nt (black), 19~nt (gray), 21~nt (blue), 23~nt (orange). All linkers are symmetric and $(n-1)/2$ bases of each linker are complementary to each grafted strand. (b) Measurements of the melting temperature as a function of concentration for the 21-nt linker for four different grafting densities: 2000 (circles), 500 (squares), 100 (upward triangles), and 20 (stars) DNA strands per $\mu$m$^2$. The grafting densities of A and B are equal to one another. The re-entrant concentrations are shown as points intersecting the linker-concentration axis and have a precision of roughly a factor of two. Lines in (a) and (b) are guides to the eye.}
\label{fig:density-length}
\end{figure}

To explore these two features more fully, we perform similar experiments for different grafting densities and linker lengths (i.e. binding affinities). In both cases we find qualitatively similar phase behavior with respect to increasing linker concentration as before---the melting temperature first increases and then decreases rapidly---but the melting temperatures and boundaries between regions I, II, and III change (Fig.~\ref{fig:density-length}). 

We find that increasing the linker affinity increases the melting temperature within region II, but does not affect the re-entrant concentration appreciably. We measure the melting temperature as a function of linker concentration for four linkers having different lengths: 17, 19, 21, and 23 nucleotides. We find that the melting temperature increases monotonically with increasing linker affinity, changing by about 20 degrees Centigrade between 17 nt and 23 nt (Fig.~\ref{fig:density-length}a). The re-entrant concentration, however, remains unchanged: all four linkers fail to aggregate above linker concentrations of roughly 200--300~$\mu$M (Fig.~\ref{fig:density-length}a).

Changing the grafting density, in contrast, has two effects: Both the melting temperature and the re-entrant concentration decrease with decreasing grafting density. Here we prepare different batches of colloids $A$ and $B$ with grafting densities ranging from roughly 20--2000~DNA/$\mu$m$^2$ and measure their melting temperatures as a function of linker concentration (Fig.~\ref{fig:density-length}b). As grafting density decreases, we find that the melting temperature decreases monotonically by roughly 15~degrees Centigrade over the range we explore. We also find that the re-entrant concentration decreases with decreasing grafting density, shifting from roughly 300~$\mu$M to 90~nM (about a factor of $10^4$) over the 100-fold change in grafting density explored in our experiment, hinting at a squared dependence between the re-entrant concentration and grafting density. 

Returning to our two observations concerning self-assembly from above, these experiments demonstrate that we can further increase the range over which we tune the interaction strength by adjusting the linker affinity, in addition to the linker concentration. We also find that we can adjust the range of workable linker concentrations by adjusting the grafting densities: higher grafting densities yield a wider dynamic range. Moreover, the the qualitative trends relating affinity and grafting density to the melting temperature and re-entrant concentration are consistent with our molecular-scale description of the phase behavior: Increasing linker affinity or grafting density (i.e. the concentrations of $A$ and $B$) should shift the equilibrium toward the bridged conformation $ALB$, stabilizing the solid phase and increasing the melting temperature.

\subsection*{Mean-field theory}
\label{Sec:Model}

To confirm our physical picture from above and to develop a quantitative link relating the experimental parameters to the effective interactions that emerge between colloids, we develop a mean-field theory of linker-mediated binding. Modeling the phase behavior involves two steps: 1) first we develop a model to relate the sequence, concentration, and grafting density to the multivalent free energy per particle; and 2) we relate that free energy per particle to the phase behavior of the system as a whole. 

We adapt a recent approach developed in Ref.\ \cite{di2016communication}, which models interactions between colloidal particles that result from multimeric ligand-receptor complexes to calculate the free energy per particle in the fluid and solid phases. Briefly, this involves computing the equilibrium densities of the different molecular species---bridges, half-bridges ($AL$ or $BL$), and unbound strands---and relating those densities to the free energy of multivalent binding between two particles \cite{di2016communication}. In our specific case, we compute the surface densities of half bridges $\rho_1$ and full bridges $\rho_b$ from equations of local chemical equilibrium \cite{Rogers_2011_PNAS,Varilly_2012_JCP}, starting from the densities of unhybridized strands $A$ and $B$ ($\rho_A$ and $\rho_B$), and the concentration of free linkers ($C_l$): 
\begin{eqnarray*}
\rho_1(h ) &=& \rho_A(h) {  C_l e^{-\beta \Delta G_0} \over C^\circ} \\
\rho_b( h) &=& \rho_A(h) \rho_B(h) {C_l \over C^\circ} e^{-2 \beta \Delta G_0} K( h)\\
\rho_A(h) &=& \rho_B(h)=\Psi-\rho_1(h)-\rho_b(h),
\label{eq:rhoA}
\end{eqnarray*}
where $\Psi$ is the surface density of grafted strands, $\beta = 1/k_BT$ is the reciprocal of the thermal energy $k_BT$, $\Delta G_0$ the hybridization free energy of pairing $A$ with $L$ or $B$ with $L$, $C^\circ=1$~M is the reference concentration at which $\Delta G_0$ is defined, and $K(h)$ is an effective area accounting for configurational costs associated with bridge  \cite{Mognetti_2012_SoftMatter,Varilly_2012_JCP}. For simplicity, we assume that the hybridization free energies of reacting $A$ with $L$ and $B$ with $L$ are the same. 

Because the grafted molecules are immobile and the densities are spatially nonuniform in the contact region between two particles, we evaluate the equilibrium densities locally. We define $h$ to be the local distance between the particles' surfaces: $h$ is a minimum along the line of centers and extends to the maximum separation distance at which bridges can form. Since $\rho_l(h)$ and $\rho_b(h)$ are coupled for all values of $h$ by the concentration of free linkers $C_l= C^0_l - \rho [ n_1+ {n_b \over 2}]$, where $\rho$ is the density of colloids, the numbers of half-bridges $n_1$ and bridges $n_b$ per particle are calculated self-consistently by integrating $\rho_1$ and $\rho_b$ over the particles' surfaces. See SI Section~S1 for details. 

Next we derive the phase boundaries by equating the chemical potential of the fluid to the chemical potential of the solid. We model the solid phase as a cluster of particles with coordination number $Z$. As is usually done for particles whose interaction range $\delta$ is much smaller than their radius ($\delta \ll R$), we use a cell model in which particles in the solid phase are assumed to move independently within a volume $v_f$ \cite{sear1999stability,charbonneau2007gas}. Following the same arguments as Ref.\ \cite{charbonneau2007gas}, 
we find that the melting temperature is given by
$\Delta f_\mathrm{coll} / k_B T_m  = \log \left( \rho v_f \right) + 1 $, where $\Delta f_\mathrm{coll}$ is the free energy per particle. It should be noted that the previous expression has been derived using pairwise square-well potentials with an attractive well-depth equal to $\Delta f_\mathrm{coll}$ and interaction range equal to $\delta$ with $v_f=(\delta/2)^3$ \cite{sear1999stability}. In the present case, $\Delta f_\mathrm{coll}$ is not strictly pairwise since the concentration of free linkers $C_l$ depends on the distances between all interacting particles. However, we ignore this effect. 

\begin{figure}[h]
\includegraphics[width=0.98\columnwidth]{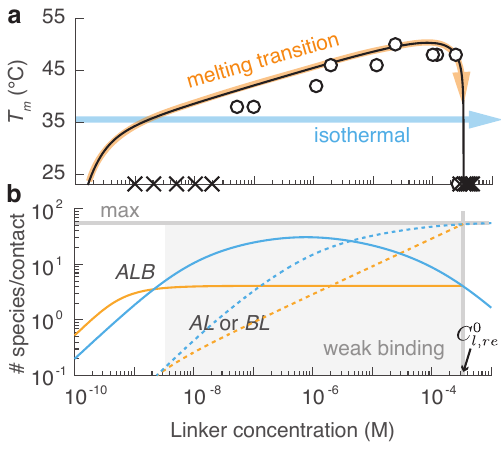}
\caption{Predictions of the melting temperature $T_m$ and number of bound molecular species per contact between particles versus linker concentration. (a) The full theory predicts a phase diagram that matches experimental measurements (points) quantitatively above linker concentrations of roughly 10~nM. (b) Predictions of the number and type of bound molecular species---either bridges (solid lines) or half-bridges (dashed lines)---between a pair of interacting particles help explain our observations. Orange curves show the number of molecular species at the melting transition (the orange path in (a)); blue curves show the number of bridge and half-bridge species at a fixed temperature: $T=35^\circ$C (the blue path in (a)). Linker concentrations at which the number of bridges at the melting transition is constant correspond to the weak-binding limit. The horizontal gray line shows the maximum number of possible bound species; the vertical gray line shows the re-entrant concentration.}    
\label{fig:theory-compare}
\end{figure}

Our mean-field theory reproduces the phase diagram that we find in our experiments. Figure~\ref{fig:theory-compare}a shows a comparison between our experimentally measured melting temperatures and the predictions of our mean-field theory for the 19-nt linker and a grafting density of 2000~DNA/$\mu$m$^2$. Both agree quantitatively above linker concentrations of roughly 10~nM: the melting temperature increases logarithmically from 40--50 degrees Centigrade upon increasing linker concentration up until roughly 300~$\mu$M, at which point the melting temperature plummets. We note a discrepancy between theory and experiment at the lowest linker concentrations ($<10$~nM). Here we find suspensions that are always disaggregated in experiment, whereas our theory predicts a melting temperature that decreases more slowly. We hypothesize that this disagreement is due to kinetic limitations and a difficulty for our system to equilibrate on experimental time scales.

Examining the number of multimeric complexes predicted by our mean-field theory confirms our molecular-scale description of the nature of the transitions between regions I, II, and III. Figure~\ref{fig:theory-compare}b shows the number of bridges and half-bridges ($AL$ or $BL$) in the contact region between two particles within the solid phase as a function of linker concentration for two distinct pathways through the phase diagram. Following a path at constant temperature (indicated by the blue arrow in Fig.~\ref{fig:theory-compare}a), we find that the number of bridges in equilibrium is $<1$ per contact at linker concentrations below roughly 1~nM: there are too few bridges to stabilize the solid phase. Upon increasing linker concentration, the number of bridges $ALB$ increases monotonically until linker concentrations of roughly 1~$\mu$M. Within this intermediate region, bridges greatly outnumber half-bridges. As the linker concentration is increased further, the number of bridges decreases as half-bridges take over and eventually saturate nearly all grafted molecules at the re-entrant concentration. 

Taking an alternative path through the phase diagram, one which follows the phase boundary, highlights another unique feature of our system: the number of bridges at the melting transition is roughly constant above a certain linker concentration (again about $>10$~nM). Following the path indicated by the orange arrow in Figure~\ref{fig:theory-compare}a, our mean-field theory predicts that there are roughly four bridges per contact for all linker concentrations above 10~nM and below the re-entrant concentration (Fig.~\ref{fig:theory-compare}b). This prediction hints at the possibility that our experiments and theory occur in the so-called ``weak binding limit'', in which the free energy per particle is approximated by the average number of bridges multiplied by the thermal energy, $\Delta f_\mathrm{coll} \approx -\langle n_b \rangle k_B T$ \cite{Biancaniello_2005_PRL,Rogers_2011_PNAS,Rogers_2012_PNAS}. 

\subsubsection*{Weak-binding limit: Melting temperature}
We explore the possibility that our experiments might be described by a weak binding limit approximation. Specifically, we take the limit of our full theory as $\Delta G_0 \rightarrow \infty$, resulting in a compact analytic expression for $T_m$: 
\begin{equation}
\frac{\Delta H_0}{k_B T_m} - \frac{\Delta S_0}{k_B} = \ln{\left[\frac{Z\pi R v_0}{2(\ln{(\rho v_f)}+1)}\frac{\Psi_A \Psi_B C_l^0}{C^\circ}\right]},
\label{eqn:weakbindinglimit}
\end{equation}
where $\Delta H_0$ is the enthalpy change of hybridization, $\Delta S_0$ is the entropy change of hybridization, $R$ is the particle radius, $\Psi_i$ is the grafting density of particle species $i$, and $v_0$ is a microscopic interaction volume defined as $v_0 = \int K(h)dh$ (see SI Section~S2 for details). Examining Equation  \ref{eqn:weakbindinglimit}, we find that the free energy of hybridization at the melting transition $\Delta G_0(T_m) = \Delta H_0 - T_m\Delta S_0$, which depends on the linker sequence, is balanced by an entropic term, which has a logarithmic dependence on the grafting densities and linker concentration---the other two independent variables in our experiment. 

\begin{figure}
\includegraphics[width=0.98\columnwidth]{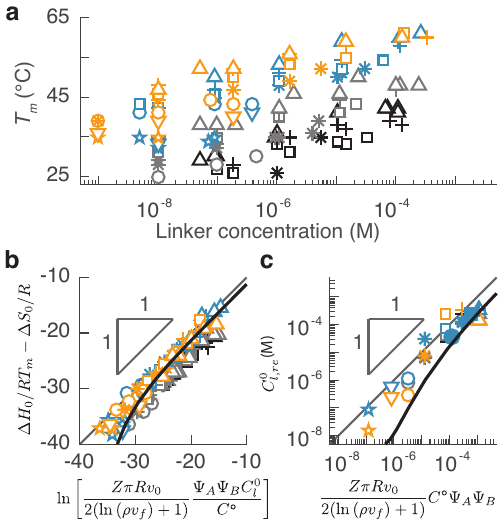}
\caption{Melting temperatures and re-entrant concentrations collapse onto master curves.  (a) shows roughly 200 measurements of the melting temperature (points) for different combinations of linker concentration, linker length, and grafting density. Color indicates linker length: 17~nt (black), 19~nt (gray), 21~nt (blue), and 23~nt (orange); and symbols indicate grafting density: 2000 (upward triangle), 1000 (plus), 500 (square), 200 (asterisk), 100 (circle),  50 (downward triangle), and 20 (star) DNA strands per $\mu$m$^2$. (b) The same data collapse when rescaled according to Equation~\ref{eqn:weakbindinglimit}. (c) The measured re-entrant concentrations collapse and exhibit a power-law dependence on the grafting density $\Psi$, as given by Equation~\ref{eqn:highconc}. Filled symbols correspond to unequal grafting densities. The solid black curves in (b-c) show predictions of the full theory.    
\label{fig:collapse}}
\end{figure}

To test the predictions of our weak-binding limit model, we measure the melting temperatures for hundreds of unique combinations of linker lengths, linker concentrations, and particle grafting densities for the case where $\Psi_A = \Psi_B$.  The data that we find shows melting temperatures ranging from roughly 25--65 degrees Centigrade, re-entrant concentrations spanning from 1~nM--300~$\mu$M, and follows the same basic trends as before: melting temperature decreases with decreasing linker concentration, decreasing grafting density, and decreasing linker length, whereas the re-entrant concentration only decreases with decreasing grafting density (see Fig.~\ref{fig:collapse}a).

Remarkably, we find that all 200+ measurements of the melting temperature collapse to a single master curve when rescaled according to Equation~\ref{eqn:weakbindinglimit}.  Figure~\ref{fig:collapse}b shows the hybridization free energy evaluated at the melting temperature versus the right-hand side of Equation~\ref{eqn:weakbindinglimit}, as well as the predictions from the weak-binding-limit model (the $y=x$ line) and the full theory. We find that all data collapse to a narrow band which falls just below the weak-binding predictions. The spread in the rescaled data reflects a variation in the melting temperatures of roughly $\pm1.5$ degrees Centigrade, which is consistent with our experimental precision of roughly $1^\circ$C. The offset indicates a minor discrepancy between the weak-binding model and our experimental measurements of about 3 degrees Centigrade. We note that the full theory is also offset slightly from the weak-binding limit, though by a smaller extent, and that the offset grows at the lowest grafting densities and linker concentrations explored. Here we would expect the weak binding approximation to break down since the depletion of free linkers---an effect not considered in the weak binding limit---becomes important.

\subsubsection*{Weak-binding limit: Re-entrant concentration}
Next we explore the dependence of the re-entrant concentration on the grafting density in the weak binding limit. Specifically, we take the limit of our full theory when both the binding is weak and the concentration of linkers is large. Here we find the following expression (see SI Section~S2 for details):
\begin{equation}
C_{l,re}^0 = \frac{Z\pi R v_0}{2(\ln{(\rho v_f)}+1)}C^\circ \Psi_A \Psi_B.
\label{eqn:highconc}
\end{equation}

We see immediately that this expression for the re-entrant concentration makes two important predictions: (1) it does not depend on the hybridization free energy and thus does not depend on the linker sequence; and (2) it scales as the product of the grafting densities of particles $A$ and $B$: higher grafting densities yield higher re-entrant concentrations. Both of these features are consistent with our observations from before (see Fig.~\ref{fig:density-length}), and the scaling of the re-entrant concentration with $\Psi_A\Psi_B$ is reminiscent of the squared dependence we saw previously.

To test the predictions of Equation~\ref{eqn:highconc} we perform two types of experiments: one in which we decrease the grafting densities of both particle species together from 2000--20~DNA/$\mu$m$^2$ for each of the four linkers, and another in which we hold the grafting density of particle A at 2000~DNA/$\mu$m$^2$ and decrease the grafting density of particle B for a single linker (19 nt). In both cases we measure the re-entrant concentration to a precision of roughly a factor of two.

We find that our measurements of the re-entrant concentration again collapse well when plotted against the predictions of Equation~\ref{eqn:highconc}, confirming the dependence on the grafting densities. In both cases---either equal grafting densities or mixed grafting densities---we observe a scaling of the re-entrant concentration that goes as the product of the two densities over a range spanning roughly four orders of magnitude (Fig.~\ref{fig:collapse}c). The fact that our data collapses so well confirms the prediction that the re-entrant concentration depends not on the grafting density alone, but on the product of the grafting densities of the two particles. Furthermore, these predictions tell us directly the maximum linker concentration that we can use for a self-assembly experiment: it must be less than the re-entrant concentration in order for the particles to assemble. The predictions of the full theory show a similar scaling for grafting densities above 50 strands$/\mu$m$^2$.

The weak-binding limit predictions of the melting temperatures and re-entrant concentrations (Eqs. \ref{eqn:weakbindinglimit} and \ref{eqn:highconc}) agree quantitatively with our experimental measurements and provide indispensable tools for programming self-assembly. Specifically, they provide simple, closed-form analytic expressions that predict the melting temperature and re-entrant concentration from the experimental inputs: linker sequence, linker concentration, and grafting density. Returning to our original motivation of fully addressible self-assembly, Equation~\ref{eqn:weakbindinglimit} can be used to choose combinations of linker sequences and concentrations that would match the melting temperatures (and thus binding affinities) of dozens of pairs of interacting particles; Equation~\ref{eqn:highconc} can be used to prepare DNA-coated colloids with sufficiently high grafting densities such that the re-entrant concentration is higher than the intended linker concentrations.

\subsection*{Combining multiple linkers}

\begin{figure*}
\onecolumngrid
\centering
\includegraphics[width=1.5\columnwidth]{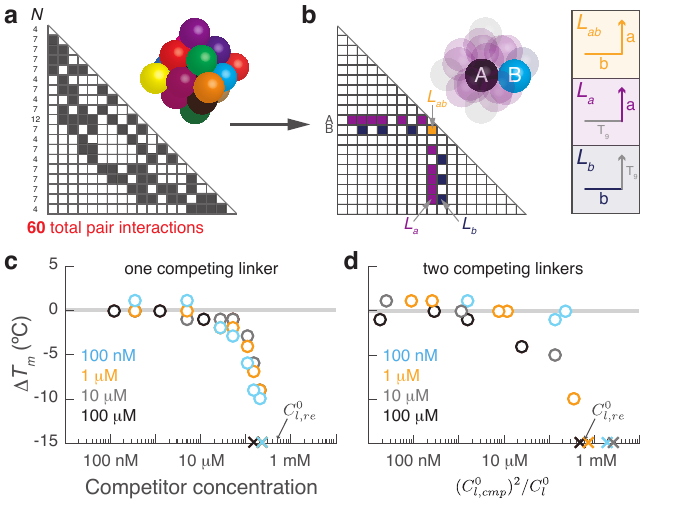}
\caption{ Effects of multiple competing linker sequences on pair interactions. (a) Fully addressable self-assembly of a 19-particle dipyramid requires specifying 60 independent pair interactions, shown in the interaction matrix, with some particles needing up to 12. The number of specific interactions per particle $N$ is shown to the left of the matrix. (b) We test the feasibility of this design by devising a simplified experimental system consisting of two particle species, $A$ and $B$, and three linker sequences $L_{ab}$, $L_a$, and $L_b$. The monovalent linkers $L_a$ and $L_b$ compete with the divalent linker $L_{ab}$, and simulate the role of other linkers binding A and B to other particle species. (c) Measurements of the change in melting temperature of particles A and B for increasing concentrations of a single competing linker $L_a$ (points). Colors indicate different concentrations of $L_{ab}$: 100~$\mu$M (black), 10~$\mu$M (grey), 1~$\mu$M (orange), and 100~nM (blue). (d) Measurements of the change in melting temperature for two competing linkers with $C_{La}^0=C_{Lb}^0$ versus $(C_{La}^0)^2/C_l^0$. x's in (c-d) show concentrations at which the particles fail to aggregate. All particles have a surface density of 2000~DNA/$\mu\textrm{m}^2$ and the re-entrant concentration is indicated by the arrow.}\label{fig:sterile}
\twocolumngrid
\end{figure*}

Thus far we have considered a situation in which two particle species interact with one another via a single linker sequence. However, we ultimately aim to use combinations of many linker sequences to specify the complex interaction matrix between many different particle species simultaneously. For instance, to self-assemble a modest structure formed from only a few same-size particles---like the 19-particle dipyramid shown in Figure~\ref{fig:sterile}a---requires programming $>50$ specific interactions.  Within the framework we propose, each particle species must then bind to multiple unique linker sequences---one for each pair interaction---which requires that some individual particles will need as many as 12 different linker sequences to specify their interactions with their neighbors in the final structure. This begs the question: Does adding multiple linker sequences, some of which bind to the same particle species and thus the same grafted sequence, interfere with binding or compromise the validity of our models developed above? For instance, does adding a linker sequence which binds A to C interfere with the binding of a linker which binds A to B?

We investigate the crosstalk between different linker sequences by designing a slight variant on our previously described  experiments. Specifically, we modify our 19-nt linker so that it only binds to one grafted strand and replace the other half of the linker's bases with a series of inert thymines. Thus, instead of forming bridges, our modified linkers can only form half-bridges, which inhibit assembly. We call these modified linkers ``competitors.'' In the context of complex self-assembly, the competitors act like other linker species that would bind particles A or B to other particle types in the same solution (Fig.~\ref{fig:sterile}b).  Indeed, we design two of these: $L_a$ which binds to particle $A$ but not particle $B$; and $L_b$ which binds to particle $B$ but not $A$. We then mix these competitors together with our active linker $L_{ab}$---the only linker sequence that can form bridges---and investigate the effect of increasing competitor concentrations on the melting temperature of our colloidal suspension. 

If the thermodynamics of linker-mediated binding were unaffected by the presence of other linkers in solution, we would not expect to observe a change in the melting temperature upon the addition of the competitors. Figure~\ref{fig:sterile}c shows the change in melting temperature of the binary mixture of colloids $A$ and $B$ upon addition of a single competitor species $L_a$ at different concentrations of the active linkers. We find that up to quite high concentrations of the competitor, the melting temperature is independent of the amount of competitor added (see Fig.~\ref{fig:sterile}c), even when the competitor is added in 100-fold excess of the active linker (see for example 100~nM active linker and 10~$\mu$M competitor). Indeed, up to competitor concentrations of roughly 10~$\mu$M we find that the melting temperature is constant for active linker concentrations ranging from 100~nM--100~$\mu$M. Above this concentration, the melting temperature decreases and the particles eventually melt by competitor concentrations of a few hundred $\mu$M. This behavior is independent of the amount of active linker $L_{ab}$ added to the suspension: we explore 100~nM, 1~$\mu$M, 10~$\mu$M, and 100~$\mu$M and find the same behavior in all cases.

A more stringent test of the competition between different linker species is to add both competitor linkers, $L_a$ and $L_b$, simultaneously. Here we find similar qualitative behavior: the melting temperature remains unchanged at low competitor concentrations and then decreases above some threshold until the particles no longer aggregate. Figure~\ref{fig:sterile}d shows the change in melting temperature versus increasing competitor concentration, rescaled by $(C_{l,cmp}^0)^2/C_l^0$, where $C_{l,cmp}^0$ is the concentration of $L_a$ and $L_b$ (we keep the concentrations of both competitors the same). Below a rescaled concentration of roughly 10~$\mu$M, the melting temperature does not change. At higher concentrations the melting temperature decreases with increasing competitor concentration, with a transition that appears to depend weakly on the active linker concentration: the melting temperature of the highest active linker concentration (100~$\mu$M) decreases by roughly 5~degrees Centigrade at a scaled competitor concentration of about 10~$\mu$M, whereas the melting temperature of the 100-nM active linker case does not decrease until the scaled competitor concentration exceeds $>100$~$\mu$M.

At first blush, the results we find---that adding 10--100 fold excess competitor does not interfere with the assembly of $A$ to $B$ via linker $L_{ab}$---are counterintuitive. However, the predictions of our model provide an explanation. The change in the phase behavior that we observe in our competitor experiments can be understood as a competition between half-bridges, which could be formed by either active linkers or competitors, and bridges, which can only be formed by active linkers. The observation that the melting temperature is unchanged by modest concentrations of competitor suggests that bridges are considerably more thermodynamically stable than half-bridges. Indeed, returning to Figure~\ref{fig:theory-compare}b we see that bridges outnumber half-bridges up until linker concentrations of roughly 1~$\mu$M and that half-bridges do not occupy a majority of grafted sites until concentrations above 10~$\mu$M. In other words, at low competitor or linker concentrations the system would prefer to form bridges instead of half-bridges, again due to the entropy of the free linkers dispersed in solution (see SI Section~S2). At higher concentrations, the situation changes and half-bridges become more thermodynamically stable than bridges. This same mechanism is responsible for the decrease in melting temperature that we observe in our competitor experiments, as well as for the re-entrant melting transition that we observed previously.

Most importantly, our findings demonstrate that multiple linker species can in fact be added together without interfering with one another, provided that the concentrations of linkers are below a threshold value. A back of the envelope estimation shows that linker-prescribed assembly of modest aperiodic structures can be accomplished using our scheme. Given that same-sized spherical particles are able to have at most 12 neighbors \cite{hales2005proof}, there need not be more than 12 linker sequences that bind to the same particle type, corresponding to a competitor/linker ratio of 11/1. Our experimental data show that even a 100/1 ratio of competitor to linker does not change the melting temperature appreciably for  linker concentrations in the range of 10~nM to $10~\mu$M. Thus we conclude that we could indeed prescribe the 60 total pair interactions needed to encode the 19-particle dipyramid in Figure~\ref{fig:sterile} using 60 unique linker sequences, and then match all of their pair-interaction free energies using the theory we develop above. 

\section*{Conclusions}
In this work, we have shown that linker-mediated self-assembly has a number of interesting features and distinct advantages compared to self-assembly of DNA-coated colloids due to direct hybridization. First, many distinct linker sequences can be combined to specify and tune the hundreds of specific interactions needed to encode a prescribed, aperiodic structure as the only ground state in a complex mixture of same-sized colloids. Unlike DNA interactions due to direct binding, in which every orthogonal pair interaction must be specified by a different grafted sequence and the mutual interaction strengths are hardwired once the particles are synthesize, linker-mediated interactions can be tuned \emph{in situ} by adjusting the linker concentrations. Furthermore, many specific interactions can be encoded between particles which are each grafted with a single sequence by creating cocktails of many linker sequences in the same solution: one linker sequence per pair interaction. Therefore, encoding every possible pair interaction between $P$ particle species in a linker-based system would require only $P$ distinct grafted sequences, whereas specifying the same number of interactions would require $P(P+1)/2$ grafted sequences in the direct-binding case. 

This enhanced flexibility in linker-mediated binding results from additional degrees of freedom introduced to the system---the molar concentrations of each linker sequence---which modify the interaction free energy per particle in nontrivial ways. Importantly, we show that the influence of these new degrees of freedom can be modeled quantitatively using our mean-field theory, and can thus be programmed \emph{a priori}. We stress that one technical hurdle to assembling prescribed structures from uniformly coated spheres remains: the structures must be assembled in systems containing only one of each particle species. Microfluidics-based methods have been developed to conduct experiments within such constraints \cite{lee2014synchronized}, but these considerations are beyond the scope of this article.

Second, the phase behavior of linker-mediated self-assembly is qualitatively different. Whereas the interactions between DNA-coated colloids due to direct hybridization increase monotonically with increasing DNA density, interactions between DNA-coated particles due to linker sequences dissolved in solution are non-monotonic: they first increase and then decrease upon increasing linker concentration, inducing a re-entrant melting transition in the phase diagram. Since this re-entrant transition is reproduced by our mean-field theory, which assumes local equilibrium at the molecular scale, we emphasize that the re-entrant melting transition should be generic to systems in which assembly is due to weak, multivalent binding mediated by free molecules in solution. Indeed qualitatively similar behavior is observed in a wide range of experimental systems, ranging from `squelching' in gene expression \cite{smith1996creb} to re-entrant condensation in proteins \cite{zhang2008reentrant} and nucleic acids \cite{nguyen2000reentrant} to self-assembly of virus particles \cite{asor2017crystallization}. Thus our model may find applications in a number of other  settings.

Finally, while the current study examines the phase behavior that emerges in equilibrium, we highlight that linker-based systems could also be used to study non-equilibrium routes to self-assembly. For instance, our demonstration that linker-mediated phase behavior results from the local equilibrium of molecular-scale reactions opens the door to inclusion of complex DNA-based circuits and devices from DNA nanotechnology into colloidal self-assembly, such as catalytic amplifiers, cascaded circuits, and logic gates \cite{zhang2011dynamic}. The integration of such non-equilibrium devices, which break detailed balance, could yield schemes for error correction, adaptation, and other strategies exploited by biological systems to engineer an astonishing diversity of self-assembling materials \cite{hopfield1974kinetic,hartwell1999molecular}. 

\begin{acknowledgments}
We thank Rees Garmann, Vinothan Manoharan, and Michael Hagan for fruitful discussions, and Michael Perlow and Kyra Hamel for their assistance in measuring the hybridization free energies. JL, GNP, and WBR acknowledge support from the National Science Foundation (NSF DMR-1710112) and the Brandeis MRSEC (NSF DMR-1420382). The work of BO and BMM is supported by an ARC (ULB) grant of the {\em F\'ed\'eration Wallonie-Bruxelles } and by the F.R.S.-FNRS under grant n$^\circ$ MIS F.4534.17.
\end{acknowledgments}

\bibliography{1_manuscript.bbl}

\end{document}



\title{Supporting Information: Linker-mediated phase behavior of DNA-coated colloids}

\author{Janna Lowensohn} 
\affiliation{Martin A. Fisher School of Physics, Brandeis University, Waltham, MA 02453, USA} 

\author{Bernardo Oyarz\'un} 
\affiliation{Universit\'e Libre de Bruxelles, Interdisciplinary Center for Nonlinear Phenomena and Complex Systems, Campus Plaine, Code Postal 231, Blvd du Triomphe, B-1050 Brussels, Belgium} 

\author{Guillermo Narvaez Paliza} 
\affiliation{Martin A. Fisher School of Physics, Brandeis University, Waltham, MA 02453, USA} 


\author{Bortolo M. Mognetti}
\affiliation{Universit\'e Libre de Bruxelles, Interdisciplinary Center for Nonlinear Phenomena and Complex Systems, Campus Plaine, Code Postal 231, Blvd du Triomphe, B-1050 Brussels, Belgium} 

\author{W. Benjamin Rogers} 
\email{wrogers@brandeis.edu}
\affiliation{Martin A. Fisher School of Physics, Brandeis University, Waltham, MA 02453, USA}

\date{\today}

\maketitle


\section{Experimental materials and methods}
\subsection{DNA colloidal grafting}
We synthesize DNA-grafted colloidal particles using a technique that physically grafts DNA-conjugated block copolymers to the surface of 1-micrometer-diameter polystyrene microspheres (Invitrogen)~\cite{Kim_2005_JACS}.  Briefly, the terminal hydroxyl ends of a poly(ethylene oxide)-poly(propylene oxide)-poly(ethylene oxide) triblock copolymer (Pluronic F108; BASF) are activated by p-nitrophenyl chloroformate (Sigma-Aldrich). A subsequent reaction with 5'-amino-C6-modified single-stranded DNA (ssDNA) oligonucleotides (Integrated DNA Technologies, Inc.) forms a stable carbamate linkage between F108 and ssDNA. The DNA-conjugated copolymers are then adsorbed to the surface of polystyrene microspheres (Invitrogen) in 10 mM citric acid buffer (pH = 4), and physically grafted by swelling and deswelling the polystyrene cores with toluene. We use a volume of toluene equal to the total volume of polystyrene.  Finally, DNA-grafted particles are washed and stored in an aqueous buffer containing 10 mM Tris and 1 mM EDTA (pH = 8) at a colloidal particle volume fraction of roughly 1\% (v/v). We estimate our maximum DNA density to be 6,500 DNA strands per particle, and use this value in all calculations presented. We adjust the labeling density from roughly 65--6500 strands per particle by diluting the DNA-functionalized F108 with unlabeled F108 before swelling~\cite{Kim_2006_Langmuir}.

\subsection{DNA sequence design and synthesis}
The sequences we use are designed following a prescription described in Ref. \cite{Seeman_1990_JBSD}.  All DNA sequences are designed to minimize formation of stable secondary structures (such as hairpins) and crosstalk between non-interacting sequences by ensuring that all three-base codons and their complements are used only once.  

Grafted sequences are 65-bases-long, single-stranded, and consist of an inert poly-dT spacer and functional domain on the 3' end. The poly-dT spacer sets the range of interaction; the sticky end sequences are 11-nucleotides long and provide a unique label for each bead species.  All surface-grafted strands are purified by high-performance liquid chromatography.

Soluble linker strands range from 17 to 21 nucleotides in length.  A linker of length $n$ will have $(n-1)/2$ nucleotides complementary to each of the grafted strands' sticky ends. All linkers also have one cytosine base between the two binding domains, which acts as a flexible spacer.  All linker strands are purified by standard desalting. The specific base sequences we use are given in Table 1.

\begin{figure}
\begin{center}
\includegraphics[]{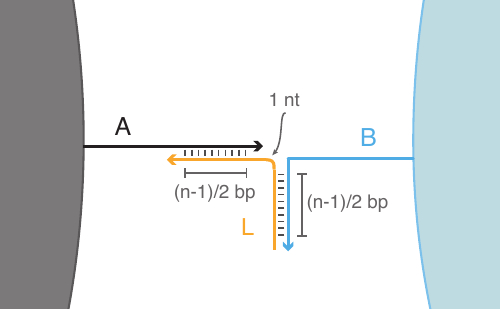}
 \end{center}
\caption{Schematic of the binding domains for grafted and soluble ssDNA strands. Grafted strands have flexible, unpaired tethers and sticky ends. Linkers are symmetric and have two binding domains of length $(n-1)/2$ separated by a single unpaired nucleotide (nt). The total length of the linker is $n$.}
\end{figure}

\begin{table}
\begin{ruledtabular}
\begin{tabular}{lll}
     Name & Modifications & Sequence\\
     \hline
     A & 5' Amino Modifier C6 & 5'-$T_{54}$ GTA TGT GGT TA-3'\\
     B & 5' Amino Modifier C6 & 5'-$T_{54}$ GAT TGA AGA GT-3'\\
     17-nt Linker $L_{ab}$ & none & 5'-ACT CTT CAC TAA CCA CA-3'\\
     19-nt Linker $L_{ab}$ & none & 5'-ACT CTT CAA CTA ACC ACA T-3'\\
     Competitor $L_a$ & none & 5'-ACT CTT CAA CTT TTT TTT T-3'\\
     Competitor $L_b$ & none & 5'-TTT TTT TTT CTA ACC ACA T-3'\\
     21-nt Linker $L_{ab}$ & none & 5'-ACT CTT CAA TCT AAC CAC ATA-3'\\
     23-nt Linker $L_{ab}$ & none & 5'-ACT CTT CAA TCC TAA CCA CAT AC-3'\\

\end{tabular}
\end{ruledtabular}
\caption {DNA sequences}.
\end{table}

\subsection{Determining thermodynamic parameters}
We measure directly the standard enthalpy change $\Delta H$ and entropy change $\Delta S$ of hybridization for all linkers using ultraviolet (UV) spectrometry, following the procedure outlined in Ref.~\cite{di2016communication}. Our experimental setup consists of a UV light source (DH-2000-BAL, Ocean Optics), a temperature-controlled cuvette holder (qpod, Quantum Northwest), and a spectrometer (Flame-S-UV-Vis-ES, Ocean Optics). Briefly, 400~microliters of sample containing 1~$\mu$M of each DNA species, 500~mM NaCl, and 1X Tris-EDTA are loaded into a quartz cuvette (Starna Cells, Inc.) and covered with a layer of mineral oil (Sigma Aldrich). The samples are heated to $90^\circ$C for 20~min and then cooled to $20^\circ$C at a rate of 0.5~$^\circ$/min while recording absorption spectra at 0.5$^\circ$C intervals. At each temperature point, the absorption spectrum is averaged over 20~s.

We use the ``baseline'' method to extract the DNA thermodynamics from our absorption measurements. Briefly, we fit the low and high-temperature baselines of the temperature-dependent absorption measured at 260 nanometers with straight lines $b_\textrm{low}(T)$ and $b_\textrm{high}(T)$ and then compute the fraction of hybridized strands $\theta(T)$ from the absorption data $A(T)$ using $\theta(T) = (b_\textrm{high}-A(T))/(b_\textrm{high}-b_\textrm{low})$. According to the two-step model of DNA hybridization, in which hybridization is represented by the bimolecular reaction $X + L \leftrightarrow XL$, the equilibrium constant $K_a = [XL]\rho_0/([X][L])$ is related to the standard free energy $\Delta G$ through $K_a = e^{-\Delta G/RT}$, where $\rho_0$ is a reference concentration, $R$ is the gas constant and $T$ is the temperature. Considering an equimolar mixture of single strands $X$ and $L$ with concentrations $\rho$, we relate the hybridized fraction $\theta(T)$ to the standard free energy via 
\begin{equation}
    \ln K_a = \frac{\Delta G}{RT} = \frac{\Delta S}{R} - \frac{\Delta H}{RT} = \ln \frac{\theta(T)\rho_0}{(1-\theta(T))^2\rho}.
\end{equation}
Thus we obtain the enthalpy change and entropy change from the slope and intercept of a plot of $\ln K_a$ versus $1/T.$

We confirm that the two-state model is accurate by performing the same experiment at different strand concentrations and inferring the enthalpy change and entropy change from an Arrhenius plot of $1/T_m$ versus $\ln \rho$, where $T_m$ is defined as the temperature at which $\theta = 0.5.$ The values obtained from both methods agree to within statistical uncertainty.

To account for the possible influence of the polyT spacers on the DNA thermodynamics, we use the full linker sequences and modified grafted sequences, which contain the last twelve nucleotides on the 3' ends.The enthalpy and entropy changes that we find are given in Table 2.

\begin{table}
\begin{ruledtabular}
\begin{tabular}{lllll}
     Name & A side && B side &\\
     & $\Delta H$ (kcal/mol) & $\Delta S$ (cal/mol$\cdot$K) &  $\Delta H$ (kcal/mol)& $\Delta S$ (cal/mol$\cdot$K)\\
     \hline
     17-nt Linker $L_{ab}$ & -55.6 & -154.6  & -52.4 & -148.0\\
     19-nt Linker $L_{ab}$ & -75.3 & -213.4 & -58.4 & -162.6\\
     21-nt Linker $L_{ab}$ & -76.6 & -215.9 & -82.1 & -228.1\\
     23-nt Linker $L_{ab}$ & -85.9 & -240.1 & -68.0 & -185.5\\
\end{tabular}
\end{ruledtabular}
\caption{DNA thermodynamics}.
\end{table}

\subsection{Measuring the melting temperature}
We measure the melting temperature of our colloidal suspensions using optical microscopy. Samples are prepared by mixing particle species A, particle species B, and free ssDNA solution in a 1:1:2 ratio.  The final solutions contain linker strands and competitor strands in 10~mM Tris/1~mM EDTA/500~mM NaCl and have particle volume fractions of roughly 0.25\% (v/v) per species.  Linker solutions and competitor solutions are prepared at various concentrations, which are measured directly using UV spetrophotometry (NanoDrop 2000c; Thermo Scientific).  The final solution of free ssDNA is created by combing these two solutions of known concentration.

Next we prepare sample chambers using two coverslips (No. 1; VWR) that are bonded togerther and sealed with silicone vacuum grease (Dow Corning) and affixed with UV-curable optical adhesive (Norland 63) or  quick-dry nail polish (big kwik dry top coat; Sally Hansen).  The coverslips are plasma cleaned for approximately 1 minute  before the chamber is assembled in order to prevent nonspecific binding of the DNA-grafted particles and the chamber walls. Finished sample chambers are roughly 1.7~$\mu$L in volume and about 70~$\mu$m in height.

We prepare a separate microscope chamber for each unique combination of linker length, linker concentration, and particle grafting density.  Once the chamber is assembled, we anneal the sample in an oven (Shell Lab, 1330FM Horizontal Airflow Oven) at $70^\circ$~C for 30 minutes and then at $45^\circ$~C for another 30 minutes.  Samples are then removed from the oven and allowed to sit at room temperature for at least 1 hour.

We image our samples on an inverted optical microscope (Nikon Eclipse TE2000-E) using a 60x oil-immersion objective (VC; Nikon).  We control the sample temperature using two heating elements: (1) we heat our sample from below using an objective heater fitted with a thermo-sensitive resistor and driven by a low-noise temperature controller (BioScience Tools); (2) we also control the sample temperature from above using a Peltier heater (TE Technology) with a hole in its center, which is bonded to the sample chamber with optical gel (Cargille Optical Gel) and controlled via software (TC 720, TE Technologies).  The Peltier is equipped with a water block.

We measure the melting temperature by slowly heating the sample while imaging it simultaneously using a digital camera. The temperature is incremented by 1 degree Centigrade, the sample is equilibrated for 5--10 minutes, and then digital images are acquired. The melting temperature is determined from this series of images by determining the lowest temperature at which $\sim 50\%$ of all particles are unbound. The singlet fraction is determined by visual inspection.  Since the melting transition is abrupt (roughly $1^\circ$C wide), we estimate an uncertainty of roughly 1$^\circ$C  associated with our measurements of the melting temperature $T_m$. Samples that are disaggregated even at room temperature and do not have a measurable melting temperature are indicated by x's in figures in the main text. 

\section{Full Theory}
\label{Sec:Full}

We consider a binary suspension of colloidal particles functionalized with DNA oligomers of type $A$ or $B$. The stoichiometry of the suspension is 1:1. $R$ denotes the particle radius and $\rho$ is the total number density of particles in the suspension.
 We define the areal density of unhybridized oligomers $A$ and $B$ by $\rho_A({\bf r})$ and $\rho_B({\bf r})$, respectively. 
$\rho_{1,X}({\bf r})$ ($X=A$ or $B$) and $\rho_b({\bf r})$ are the areal densities of half-bridges ($AL$ or $BL$) and bridges ($ALB$), respectively (see Manuscript Fig.~2). Similarly, $\rho_{s,X}$ are the areal densities of grafted oligomers bound to a single competitor sequence. 
${\bf r}$ is a bi--dimensional coordinate spanning the particle's surface.  
If $C_l$ and $C_{l,X}$ (labeled $C_{L_a}$ and $C_{L_b}$ in the main text) are the density of linkers and competitors of type $X$ not bound to any colloid, we find
\begin{eqnarray}
\rho_X({\bf r}) &=& \Psi_{X}-\rho_{1,X}({\bf r})-\rho_{s,X}({\bf r})-\rho_b({\bf r})
\nonumber \\
C_l &=&  \left[ C^0_l-\rho {n_{1,A} + n_{1,B} \over 2} - \rho { n_b \over 2} \right]
\nonumber \\
C_{l,X} &=& \left[ C^0_{l,X}- {\rho \over 2} n_{s,\overline X (X)} \right],
\label{eq:densities_free}
\end{eqnarray}
where $\Psi_{X}$ is the grafting density of oligomer $X$, while $C^0_l$ and $C^0_{l,X}$ are the total densities of linkers and competitors. We define $\overline X(X)=A$ if $X=B$ and $\overline X(X)=B$ if $X=A$. In Eq.~\ref{eq:densities_free} we have defined 
\begin{eqnarray}
n_{1,X} = \int_\Omega \mathrm{d} {\bf r} \, \rho_{1,X}({\bf r}) \,\, ,
\qquad
n_{b} = \int_\Omega \mathrm{d} {\bf r} \, \rho_{b}({\bf r}) \,\, ,
\qquad
n_{s,X} = \int_\Omega \mathrm{d} {\bf r} \, \rho_{s,X}({\bf r}) \,\, .
\label{eq:number}
\end{eqnarray}
\begin{figure}
\begin{center}
\includegraphics[scale=0.4]{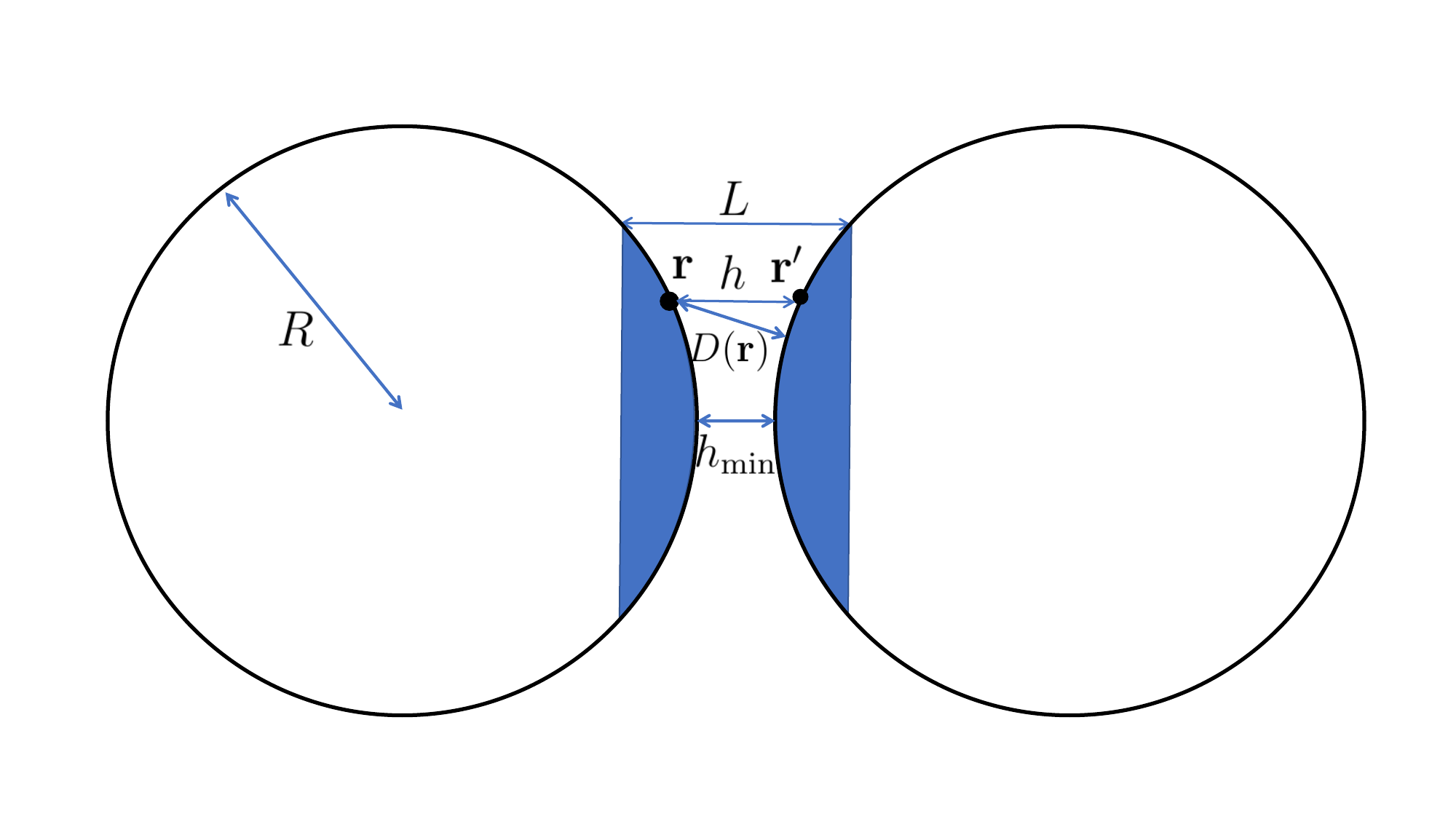}
 \end{center}
\caption{Schematic of the geometric quantities entering the Derjaguin approximation. The interacting region has been stained in blue (not drawn to scale). }\label{fig:DJ}
\end{figure}

We use a Derjaguin like approach adapted to the presence of free linkers in solution.  
In particular, we simplify the geometry of the system assuming that short oligomers tethered in the interaction region at position ${\bf r}$ cannot detect curvature effects but only see a functionalized plane placed at distance $D({\bf r})$, where $D({\bf r})$ is the distance between ${\bf r}$ and the surface of the facing colloid (see Fig.~\ref{fig:DJ}). When $R/L\gg 1$ (where $L$ is the range of interaction comparable with the length of the stretched oligomers), $D({\bf r})$ becomes comparable with the distance ($h$) between the point ${\bf r}$ and its image ${\bf r}'$ on the facing colloid (see Fig.~\ref{fig:DJ}). In view of the cylindrical symmetry of the system, $\rho_X$, $\rho_{1,X}$, $\rho_{s,X}$, and $\rho_b$ are only function of $h$ with $h_\mathrm{min}\leq h\leq L$ (see Fig.~\ref{fig:DJ}). The densities of half-bridges and bridges can then be calculated using the following equations of chemical equilibrium:  
\begin{eqnarray}
\rho_{1,X}(h) &=& \rho_X(h) {C_l \over C^\circ} e^{-\beta \Delta G^0_X}
 \label{eq:linker}
 \\
\rho_{s,X}(h) &=& \rho_X(h)  {C_{l,\overline X(X)} \over C^\circ } e^{-\beta \Delta G^0_X}
\label{eq:sterile}
 \\
\rho_b(h) &=& \rho_{X}(h) \rho_{\overline X(X)}(h)  {C_l K(h)\over C^\circ } e^{-\beta \Delta G^0_{A}-\beta\Delta G^0_{B}},
\label{eq:bridge}
\end{eqnarray}
where $K(h)/C^\circ$ is a system specific configurational-entropy term linked to the cost of binding two tethered oligomers (see Sec.~\ref{sec:K}). $\Delta G^0_A$ and $\Delta G^0_B$ are the free-solution hybridization free energies of the sticky ends on oligomers $A$ and $B$, respectively (in the main text we take them to be equal and define $\Delta G_0=\Delta G^0_A=\Delta G^0_B$). We define $C^\circ$ as the standard concentration. Eqs.~\ref{eq:linker}, \ref{eq:sterile}, and  \ref{eq:bridge} apply in the interaction region between colloids (shadowed area in Fig.~\ref{fig:DJ}). Outside the interaction region, $\rho_b=0$ (given that $K(h)=0$), while $\rho_{1,X}$ and $\rho_{s,X}$ are constant: $\rho_{1,X}=\rho_{1,X}(L)$ and $\rho_{s,X}=\rho_{s,X}(L)$. Notice that Eqs.~\ref{eq:linker}, \ref{eq:sterile}, and \ref{eq:bridge} are coupled by the concentrations of linkers $C_l$ and competitors $C_{l,X}$ at different values of $h$ (see Eqs.~\ref{eq:densities_free}).

Reference~\cite{di2016communication} provides a portable expression of the free energy of multivalent systems, which can be written as the sum of two different contributions: 
\begin{enumerate}
\item For each type of binding molecules, the free energy per particle $f_\mathrm{coll}$ includes a logarithm term that is proportional to the fraction of not-bound molecules (see the first three terms in Eq.~\ref{eq:free}). In the present system, the families of binding molecules are the competitors, the linkers, and the tethered oligomers at a given position $h$ (see Fig.~\ref{fig:DJ}). Notice that oligomers at different $h$ are treated as different molecules given that they form bridges with different affinities as specified by $K(h)$ in Eq.~\ref{eq:bridge}.
\item For each complex formed by $m$ binding molecules, $f_\mathrm{coll}$ is augmented by $(m-1) k_B T$. In the present system, dimers ($m=2$) are oligomers binding competitors or linkers (without forming a bridge), while bridges count as trimers ($m=3$).
\end{enumerate}
The free energy {\em per} particle is then given by   
\begin{eqnarray}
\beta f_\mathrm{coll} &=& \int_\Omega \mathrm{d} {\bf r} \, \sum_{X=A,B} {\Psi_X\over 2} \ln {\rho_X({\bf r}) \over \Psi_X} + {C^0_l \over \rho } \ln { C_l \over C^0_l} + \sum_{X=A,B}
{C^0_{l,X} \over \rho} \ln  { C_{l,X} \over C^0_{l,X}}
\nonumber \\
&& + \int_\Omega \mathrm{d} {\bf r} \, \left[ 
\sum_{X=A,B} {\rho_{1,X}({\bf r})+\rho_{s,X}({\bf r})  \over 2} + \rho_b ({\bf r})
\right] \,\, .
\label{eq:free}
\end{eqnarray}
The second line is equal to the sum of the number of bridges ($n_b$) and all types of half-bridges featured by the system ($n_{1,A}$, $n_{1,B}$, $n_{s,A}$, $n_{s,B}$), in which the latter is weighted by $1/2$ given that we consider an asymmetric system with $\Delta G^0_A \neq \Delta G^0_B$. Within the Derjaguin approximation, the previous expressions are simplified using 
\begin{eqnarray}
\int_\Omega \mathrm{d} {\bf r} \, \rho({\bf r}) = Z \pi \int_{h_\mathrm{min}}^L \mathrm{d} h \, \rho(h) + A_\mathrm{out} \rho(L) \, \, ,
\label{eq:DJ}
\end{eqnarray}
where $\rho$ is a density (e.g. $\rho_X$) that remains constant outside the contact region, $Z$ is the valency of the aggregate, and $A_\mathrm{out}$ the area of the region not facing any colloid. 

Equation~\ref{eq:free} can also be used to calculate the free energy of isolated colloids due to reversible binding of the linker and the competitors. For isolated particles, $\rho_{1,X}$ and $\rho_{s,X}$ are constant ($\rho_{1,X}=\rho^0_{1,X}$ and $\rho_{s,X}=\rho^0_{s,X}$) and can be calculated using Eqs.~\ref{eq:linker} and \ref{eq:sterile} with $\rho_b=n_b=0$. Defining  $f^0_\mathrm{coll}$ as the multivalent free-energy of an isolated particle in solution at a given linker concentration and temperature, the multivalent free-energy gain of moving one particle from the gas to the the solid phase is given by $\Delta f_\mathrm{coll}= f_\mathrm{coll}- f^0_\mathrm{coll}$.

\begin{figure}
\begin{center}
\includegraphics[scale=0.4]{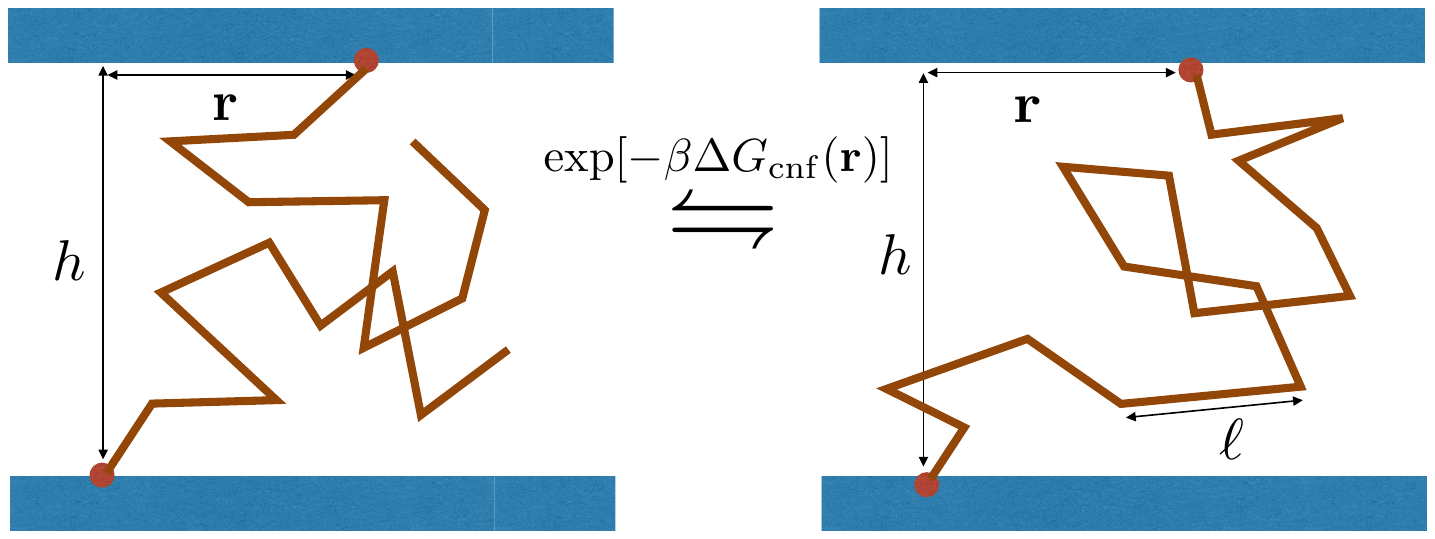}
 \end{center}
\vspace{-3.cm}
\caption{$\Delta G_\mathrm{cnf}$ accounts for the configurational volume loss of forming a bridge and is calculated using Monte Carlo simulations \cite{Varilly_2012_JCP,de2014new}.}\label{fig:Kh}
\end{figure}
\subsection{Calculation of the configurational-entropy term}\label{sec:K}

The probability ($p$) of reacting two isolated, tethered oligomers carrying complementary sticky ends  (see Fig.~\ref{fig:Kh}) is controlled by the hybridization free energy $\Delta G_\mathrm{hyb}$ ($p=\exp[-\beta \Delta G_\mathrm{hyb}]/(1+\exp[-\beta \Delta G_\mathrm{hyb}])$).
For fixed tether points, $\Delta G_\mathrm{hyb}$ is the sum of the hybridization free energy of the sticky ends free in solution ($\Delta G_0$) and a configurational-entropy term accounting for the cost of binding the ends of the two grafted oligomers to one another ($\Delta G_\mathrm{cnf}({\bf r}, h)$)

\begin{eqnarray}
\Delta G_\mathrm{hyb}({\bf r}, h) = \Delta G_0 + \Delta G_\mathrm{cnf}({\bf r}, h) \, \, .
\end{eqnarray}
Notice how $\Delta G_\mathrm{cnf}$ is function of the positions of the tethering points as well of the length and flexibility of the spacer anchoring the sticky end to the particle's surface. For rigid, ideal spacers, $\Delta G_\mathrm{cnf}$ can be calculated using theoretical arguments. For flexible spacers, Monte Carlo methods allow calculating $\Delta G_\mathrm{cnf}$ \cite{Varilly_2012_JCP,de2014new}.
Using $\Delta G_\mathrm{cnf}$, $K(h)$ reads as follows
 \begin{eqnarray}
K (h) &=& \int \mathrm{d} {\bf r}^2 e^{-\beta \Delta G_\mathrm{cnf} ({\bf r},h)} \, ,
\label{Eq:Kh}
\end{eqnarray}
where the integral is taken over the lateral displacement of the tethering points (see Fig.~\ref{fig:Kh}). Consistently with the Derjaguin approximation, we tabulate $\Delta G_\mathrm{cnf} ({\bf r},h)$ using oligomers tethered to parallel plates. We use the representation of Rogers and Crocker in which 55--base oligomers are mapped into freely-jointed chains composed of 8 Kuhn segments of length equal to 5$\,$nm \cite{Rogers_2012_PNAS}. Further details about the calculation of $K(h)$ and $\Delta G_{\rm{cnf}}$  can be found in Refs.~\cite{Varilly_2012_JCP,de2014new}.

\subsection{Numerical strategy}

In this section we illustrate the numerical strategy used to calculate the equilibrium densities of the different reacting species ($C_l$, $C_{l,X}$, $\rho_X(h)$, $\rho_{1,X}(h)$, $\rho_{s,X}(h)$, and $\rho_b(h)$), as well as the multivalent free energy ($f_\mathrm{coll}$). We use the densities of unbound monomers ($C_l$, $C_{l,X}$, and $\rho_X(h)$, see Eqs.~\ref{eq:densities_free}) as numerical variables, while the densities of dimers and trimers  ($\rho_{1,X}(h)$, $\rho_{s,X}(h)$, and $\rho_b(h)$) are constrained by Eqs.~\ref{eq:linker}, \ref{eq:sterile}, \ref{eq:bridge}. In particular, using Eqs.~\ref{eq:linker}, \ref{eq:sterile}, and \ref{eq:bridge} we can write Eqs.~\ref{eq:densities_free} as follows 
\begin{eqnarray}
\rho_X(h) &=& 
{ \Psi_{X} \over 1+ \left[ C_l + C_{l,\overline X(X) }\right] \exp[-\beta \Delta G^0_X]/C^\circ + \rho_{\overline X(X)}(h) K(h) \exp[-\beta \Delta G^0_A-\beta \Delta G^0_B]/C^\circ }
\nonumber \\
C_l &=& {C^0_l \over 1 + {\rho\over 2 C^\circ} \int_\Omega \mathrm{d}{\bf r} \sum_{X=A,B}\rho_X({\bf r}) \exp[-\beta \Delta G^0_X] + \rho_A({\bf r}) \rho_B({\bf r}) K({\bf r}) \exp[-\beta \Delta G^0_A-\beta \Delta G^0_B]}
\nonumber \\
C_{l,X} &=& { C^0_{l,X} \over 1 + {\rho \over 2 C^\circ }  \int_\Omega \mathrm{d}{\bf r} \rho_{\overline X(X)} \exp[-\beta \Delta G^0_{\overline X(X)} ] },
\label{eq:densities_itera}
\end{eqnarray}
where $\int_\Omega \mathrm{d} {\bf r} \, \cdot \rho(r)$ is defined by Eq.~\ref{eq:DJ} and $K({\bf r})=K(h)$ (see Fig.~\ref{fig:DJ}). We discretize the interval $[h_\mathrm{min},L]$ (see Fig.~\ref{fig:DJ}) using a finite set of colloid--to--colloid distances ($\{ h_i \}$) and use Eqs.~\ref{eq:densities_itera} to numerically iterate the values of $\rho_X(h_i)$, $C_l$, and $C_{l,X}$ until convergence. The free energy is then be derived from Eq.~\ref{eq:free} using Eqs.~\ref{eq:linker}, \ref{eq:sterile}, and \ref{eq:bridge}.

\section{Scaling behaviours}

In this section we derive analytic, scaling behaviours in the different regions of the phase diagram starting from the full theory presented in Sec.~\ref{Sec:Full}. In Section~\ref{Sec:HighT} we study the high temperature $T$ regime (corresponding to the $\exp[-\beta\Delta G^0_X]\to 0$ limit); Section~\ref{Sec:HighC} presents the high linker concentration $C^0_l$ limit; and Section~\ref{Sec:HighCs} presents the high competitor concentration $C^0_{l,X}$ limit. In Sec.~\ref{Sec:HighCT}, we further investigate the re-entrant behaviour of the phase diagram by considering a double scaling limit in which $C^0_l$ and $T$ are changed while keeping $C^0_l\exp[-\beta\Delta G^0_X]/C^\circ$ constant.

\subsection{High temperature limit}
\label{Sec:HighT}

Given that the hybridization free energies $\Delta G^0_X$ are increasing functions of $T$,  the high temperature limit is studied by taking the limit $\exp[-\beta \Delta G^0_X ]\sim \epsilon \to 0$, where $\epsilon$ defines the small variable used to develop the thermodynamic quantities.
Using Eq.~\ref{eq:bridge} and Eq.~\ref{eq:densities_free} we find that the density of bridges is given by
\begin{eqnarray}
\rho_{b}({\bf r}) &=& {\Psi_{A} \Psi_{B} C^0_l K({\bf r}) \over C^\circ} \exp[-\beta \Delta G^0_A-\beta \Delta G^0_B] + O(\epsilon^3) \, 
\end{eqnarray}
to leading order in $\epsilon$. A further inspection of Eqs.~\ref{eq:linker} and \ref{eq:sterile} shows that the first two leading terms (of order $\epsilon$ and $\epsilon^2$) of the half bridges in the solid phase ($\rho_{1,X}$ and $\rho_{s,X}$) are homogeneous and match the corresponding values in the fluid phase
\begin{eqnarray}
\rho_{1,X}({\bf r}) &=& \rho^0_{1,X} + O(\epsilon^3)\,\, ,
\\
\rho_{s,X}({\bf r}) &=& \rho^0_{s,X} + O(\epsilon^3)\,\, .
\end{eqnarray}
We can then expand the free energy (given by Eq.~\ref{eq:free}) at the second order in $\epsilon$ and neglect all terms related to $\rho_{1,X}$ and $\rho_{s,X}$ in the calculation of $\Delta f_\mathrm{coll}=f_\mathrm{coll}-f^0_\mathrm{coll}$, since these terms contribute equally to $f_\mathrm{coll}$ and $f^0_\mathrm{coll}$. Finally, using Eq.~\ref{eq:DJ}, we find\begin{eqnarray}
\beta \Delta f_\mathrm{coll} &=& -{n_b\over 2} =  -{Z \pi R \over 2} {\exp[-\beta \Delta G^0_A -\beta \Delta G^0_B] \Psi_{A} \Psi_{B} C^0_l \over C^\circ} v_0,
\label{eq:freeHT}
\end{eqnarray} 
where $v_0$ is a microscopic volume given by  
\begin{eqnarray}
v_0 &=&   \int_{h_\mathrm{min}}^L K(h) \mathrm{d} h
\label{eq:v0}
\end{eqnarray}
(see Eq.~\ref{eq:DJ}).

At the melting transition ($\beta_m=1/k_B T_m$) we have $-\beta_m \Delta f_\mathrm{coll}=\ln(\rho v_f)+1$ that implies (if $\Delta G^0_A=\Delta G^0_B= \Delta H_0 - T \Delta S_0$)
\begin{eqnarray}
{\Delta H_0 \over k_B T_m} - {\Delta S_0 \over k_B} = \ln \left[ 
{Z \pi R v_0 \over 2 [\ln(\rho v_f) +1]} {\Psi_A \Psi_B C^0_l \over C^\circ}
\right]\, \, .
\end{eqnarray}
Notice how the melting temperature is not affected by the presence of competitors, as corroborated by experiments at low competitor concentration.

\subsection{High linker-concentration limit }
\label{Sec:HighC}

To investigate the re--entrant transition further, we develop the full theory in the high concentration limit $1/C^0_l \to 0$. For simplicity, we do not consider competitors ($C^0_{s,X}=\rho_{s,X}=0$). From Eqs.~\ref{eq:linker} and \ref{eq:bridge} it follows that $\rho_{1,X}({\bf r}) \to \Psi_{X}$ and $\rho_{b}({\bf r}) \to 0$ in the high $C^0_l$ limit. We then consider the following expansion for the half-bridge and bridge densities 
\begin{eqnarray}
\rho_{1,X}({\bf r}) &=& \Psi_{X} + {\alpha_{1,X}({\bf r}) \over C^0_l} + {\alpha_{2,X}({\bf r}) \over (C^0_l)^2} + O\left( 1\over (C^0_l)^3 \right) 
\nonumber \\
\rho_b ({\bf r}) &=& {\beta_{1}({\bf r}) \over C^0_l} + {\beta_{2}({\bf r}) \over (C^0_l)^2} + O\left( 1\over (C^0_l)^3 \right).  
\label{eq:DevHighC}
\end{eqnarray}
An expansion of the free energy (Eq.~\ref{eq:free}) up order $1/C^0_l$ provides
\begin{eqnarray}
\beta f_\mathrm{coll} &=& \int \mathrm{d} {\bf r} \sum_X {\Psi_X \over 2} \left[
\ln \left( -{\alpha_{1,X}({\bf r}) + \beta_{1}({\bf r}) \over \Psi_X C^0_l} \right)+{1\over C^0_l} {\alpha_{2,X}({\bf r})+\beta_{2}({\bf r}) \over \alpha_{1,X}({\bf r})+\beta_{1}({\bf r})}
\right] 
\nonumber \\
&&+{n_b \over 2} - {\rho N_X\over 2 C^0_l } + O\left( {1\over (C^0_l)^2} \right),
\label{eq:freeHC1}
\end{eqnarray}
where $N_X$ is the number of oligomers tethered to particle $X$ ($N_X= 4 \pi R^2 \Psi_X$). Using Eqs.~\ref{eq:DevHighC} and Eq.~\ref{eq:linker}, we obtain 
\begin{eqnarray}
\alpha_{1,X}({\bf r}) + \beta_{1}({\bf r}) &=& -{\rho_0 \Psi_X \over \exp[-\beta \Delta G^0_X]}
\label{eq:alpha1}
\\
\alpha_{2,X}({\bf r}) + \beta_{2}({\bf r}) &=& \rho {N_A+N_B \over 2} \left( \alpha_{1,X}({\bf r}) + \beta_{1}({\bf r}) \right) -{C^\circ \over \exp[-\beta \Delta G^0_X]} \alpha_{1,X} ({\bf r})
\nonumber \\
&=&-\rho C^\circ \Psi_X {N_A+N_B \over 2 \exp[-\beta \Delta G^0_X] } -{C^\circ \over \exp[-\beta \Delta G^0_X]} \alpha_{1,X} ({\bf r}) \,\, .
\label{eq:alpha2}
\end{eqnarray}
Instead, using Eq.~\ref{eq:bridge} (along with Eq.~\ref{eq:alpha1}), we obtain
\begin{eqnarray}
\beta_1({\bf r})= C^\circ \Psi_A \Psi_B K({\bf r}) \, \, .
\label{eq:beta1}
\end{eqnarray}
Using Eqs.~\ref{eq:DevHighC}, \ref{eq:alpha1}, \ref{eq:alpha2} and \ref{eq:beta1} we simplify Eq.~\ref{eq:freeHC1} as follows
\begin{eqnarray}
\beta f_\mathrm{coll} &=& \int \mathrm{d} {\bf r} \sum_X {\Psi_X \over 2} \left[
\ln \left( {C^\circ \over  C^0_l} \exp[\beta \Delta G^0_X] \right)+ {\rho \over C^0_l} {N_A + N_B \over 2} -{C^\circ \over C^0_l} \exp[\beta \Delta G^0_X]  
\right] 
\nonumber \\
&& - {n_b \over 2} - {\rho N_X \over 2 C^0_l } + O\left( {1\over (C^0_l)^2} \right),
\end{eqnarray}
from which we derive the following expression for $\Delta f_\mathrm{coll}$  
\begin{eqnarray}
\beta \Delta f_\mathrm{coll} &=&  -{ n_b \over 2} = - {Z \pi R \Psi_A \Psi_B C^\circ v_0 \over 2 C^0_l},
\label{eq:freeHC}
\end{eqnarray}
where $v_0$ has been defined in Eq.~\ref{eq:v0}. The re-entrant concentration ($C^0_{l,re}$) is the solution of the following equation
\begin{eqnarray}
\beta \Delta f_\mathrm{coll} =\ln(\rho v_f) + 1 =   - {Z \pi R \Psi_A \Psi_B C^\circ v_0 \over 2 C^0_{l,re}} = \ln(\rho v_f) + 1 .
\label{eq:freeHC}
\end{eqnarray}

\subsection{High linker-concentration and high temperature limit }
\label{Sec:HighCT}

In this section we study the re--entrant transition in the phase diagram at high temperature by developing the full theory in the limit $\exp[-\beta \Delta G^0_X] \to 0$, $C^0_l \to \infty$ at constant $\eta_X$ ($X=A$, $B$), defined as 
\begin{eqnarray}
\eta_X &=& {C^0_l \over C^\circ } e^{- \beta \Delta G^0_X} \, .
\end{eqnarray}
As done in the previous section, we do not consider competitors. We start by  expanding Eqs.~\ref{eq:linker} and \ref{eq:bridge}  as follows (in which we define the small variable $\epsilon \sim 1/C^0_l \sim \exp[-\beta \Delta G^0_X]$)
\begin{eqnarray}
\rho_{1,X}({\bf r}) &=& {\Psi_X \eta_X \over 1+\eta_X} +\delta \rho_{1,X} ({\bf r}) + O(\epsilon^2)
\nonumber \\
\rho_b({\bf r}) &=& { \Psi_A \Psi_B K({\bf r}) \over (1+\eta_A) (1+\eta_B)}{C^0_l \over C^\circ} e^{-\beta \Delta G^0_A -\beta \Delta G^0_B} + O(\epsilon^2)
\nonumber \\
\delta \rho_{1,X}({\bf r}) &=& -{\eta_X\over 1+\eta_X } \left[ \rho_b({\bf r}) + {\eta_X \Psi_X N_X\over (1+\eta_X)^2} {\rho \over C^0_l} \right] \,\, .
\label{eq:rho1X-HCT}
\end{eqnarray}
Next we show that 
\begin{eqnarray}
{\Psi_X \over 2} \ln {\rho_X({\bf r})\over \Psi_X} = \Lambda - {\rho_{b}({\bf r}) \over 2} + O(\epsilon^2) \, \, ,
\label{eq:rhoX-HCT}
\end{eqnarray}
where $\Lambda$ is a constant term. Using Eqs.~\ref{eq:rhoX-HCT} and  \ref{eq:rho1X-HCT} it is easy to show that $\Delta f_\mathrm{coll}$ is equal to half the number of bridges $n_b$. In particular we have:
\begin{eqnarray}
\beta \Delta f_\mathrm{coll} &=& - {n_b \over 2} + O(\epsilon^2)
\nonumber \\
&=& - { Z \pi R \Psi_A \Psi_B C^0_l v_0 \over 2 C^\circ (1+\eta_A) (1+\eta_B) } \exp[-\beta\Delta G^0_A - \beta\Delta G^0_B] + O(\epsilon^2)
\nonumber
\\ &=& - { Z \pi R \Psi_A \Psi_B C^0_l  \exp[-\beta \Delta G^0_A - \beta \Delta G^0_B] v_0 \over 2 C^\circ (1+C^0_l \exp[-\beta \Delta G^0_A]/C^\circ) (1+C^0_l \exp[-\beta\Delta G^0_B]/C^\circ ) }  + O(\epsilon^2).
\nonumber \\
\label{eq:freeHTC}
\end{eqnarray}
The previous expression interpolates the two expressions of $\Delta f_\mathrm{coll}$ found in the high $T$ (see Eq.~\ref{eq:freeHT}) and in the high $C^0_l$ regime (see Eq.~\ref{eq:freeHC}). In particular, Eqs.~\ref{eq:freeHT} and \ref{eq:freeHC} are the leading terms of Eq.~\ref{eq:freeHTC} in the large $\Delta G^0_X$ limit at constant $C^0_l$ and in the large $C^0_l$ limit at constant $\Delta G^0_X$, respectively.

\begin{figure}[ht!]
\begin{center}
\includegraphics[scale=0.4]{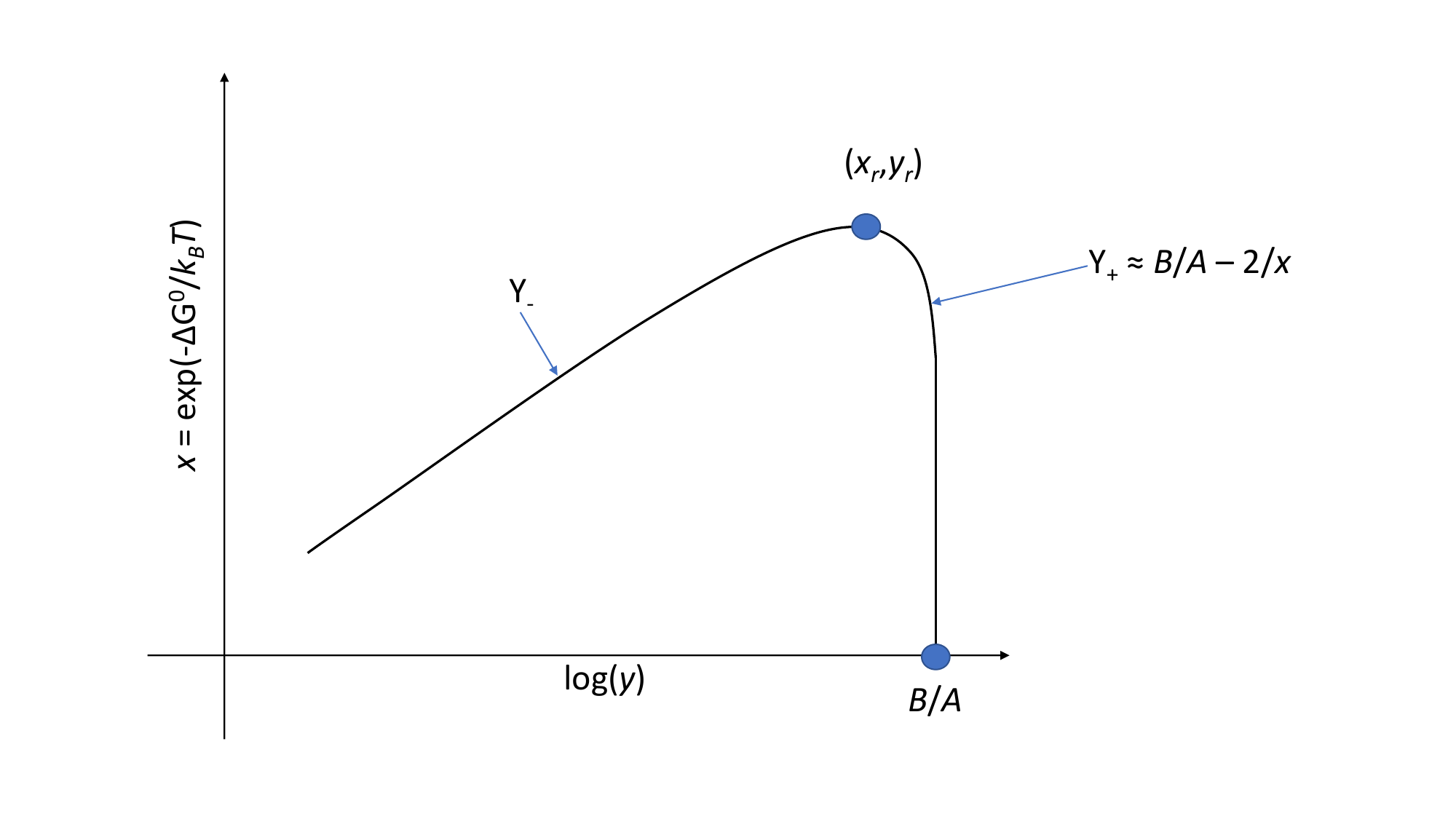}
 \end{center}
\caption{ Phase diagram in the high $T$ and high $C^0_l$ limit ($y=C^0_l/C^\circ$, see text)}\label{fig:PD}
\end{figure}
We now use Eq.~\ref{eq:freeHTC} to study the phase boundary in the high $T$, high $C^0_l$ limit, where the phase boundary is re-entrant. As illustrated in the previous section, at the transition we have $-\beta_m \Delta f_\mathrm{coll}=\ln(\rho v_f)+1$ with $\beta_m=1/T_m k_B$. Using Eq.~\ref{eq:freeHTC}, and defining $\alpha_{X}=\exp[-\beta \Delta G^0_X/k_B T_m]$ and $y=C^0_l/C^\circ$, we find 
\begin{eqnarray}
A = B {\alpha_{A} \alpha_{B} \over (1+ y \cdot \alpha_{A}) (1+ y \cdot \alpha_{B})},
\label{Eq:melting0}
\end{eqnarray}
where 
\begin{eqnarray}
A = - [ \ln (\rho v_f) +1]
& \qquad \qquad &
B = -{Z \pi R\over 2} \Psi_A  \Psi_B v_0 \, \, .
\end{eqnarray}
From Eq.~\ref{Eq:melting0} we obtain 
\begin{eqnarray}
y_\pm &=& { B \alpha_{A} \alpha_{B} -A (\alpha_{A}+\alpha_{B})  \pm \sqrt{ A^2 (\alpha_{A}-\alpha_{B})^2-2 A B \alpha_{A} \alpha_{B}(\alpha_{A}+\alpha_{B}) + B^2 \alpha_{A}^2 \alpha_{B}^2 } \over 2 A \alpha_{A} \alpha_{B} } \, .
\nonumber \\
\end{eqnarray}
$y_+$ and $y_-$ correspond to the two concentrations of the phase diagram having melting temperatures equal to $T_m$ (see Fig.~\ref{fig:PD}). The re-entrant point $\{T_{r,m}, \, y_r \}$ is determined by the relation $y_1=y_2$ and is given by
\begin{eqnarray}
x_r \equiv\exp[-\Delta G_0/k_B T_{r,m}] = {4A \over B} & \qquad \qquad & y_r ={B \over 4A},
\end{eqnarray}
assuming $\alpha_A=\alpha_B=\alpha$ (see Fig.~\ref{fig:PD}).
When $\alpha$ becomes very big (low $T$ limit), $y_+$ is approximated by 
\begin{eqnarray}
y_+ \sim {B\over A} - {2 \over \alpha} \, ,
\end{eqnarray}
implying that the corrections to the re-entrant concentration ($C^0_{l,re}=B/A$) are exponentially small at low $T$.

\subsection{High competitor-concentration limit}
\label{Sec:HighCs}

In this section we develop the full theory in the limit of high competitor concentrations $C^0_{l,S} \equiv  C^0_{l,A} = C^0_{l,B} \to \infty$. For simplicity, we consider the symmetric system with $\Psi = \Psi_A = \Psi_B$ and $\Delta G_0 = \Delta G^0_A = \Delta G^0_B$ (accordingly, $\rho_1\equiv \rho_{1,A}=\rho_{1,B}$ and $\rho_s\equiv\rho_{s,A}=\rho_{s,B}$).
Using Eqs.~\ref{eq:densities_free}, we develop Eqs.~\ref{eq:linker}, \ref{eq:sterile}, and \ref{eq:bridge} as follows 
\begin{eqnarray}
\rho_1 &=& {\alpha_1 \over C^0_{l,S} }+ {\beta_1 \over (C^0_{l,S})^2} + {\gamma_1 \over (C^0_{l,S})^3}+ O\left( 1\over (C^0_{l,S})^4 \right)
\nonumber \\
\rho_s &=& \Psi+{\alpha_s \over C^0_{l,S} }+ {\beta_s \over (C^0_{l,S})^2} + {\gamma_s \over (C^0_{l,S})^3}+ O\left( 1\over (C^0_{l,S})^4 \right)
\nonumber \\
\rho_b &=& {\alpha_b \over (C^0_{l,S})^2 }+ {\beta_b \over (C^0_{l,S})^3} + {\gamma_s \over (C^0_{l,S})^4}+ O\left( 1\over (C^0_{l,S})^5 \right)
\label{eq:den1HCls}
\end{eqnarray}
with
\begin{eqnarray}
\alpha_1 &=& C^0_l \Psi  
\nonumber \\
\alpha_2 &=& - (C^0_l)^2 \Psi- {1\over 2} C^0_l N \rho - C^0_l \Psi C^\circ \exp[\beta \Delta G_0]
\nonumber \\
\alpha_s &=& -C^0_l \Psi - \Psi C^\circ \exp[\beta \Delta G_0]
\nonumber \\
\beta_s ({\bf r}) &=& (C^0_l)^2 \Psi + {1\over 2} C^0_l N \Psi \rho- C^0_l K({\bf r}) \Psi^2 C^\circ + 2 C^0_l \Psi C^0 \exp[\beta \Delta G_0] 
\nonumber \\
&& -{1\over 2} N \Psi \rho C^\circ \exp[\beta \Delta G_0] 
 +\Psi (C^\circ)^2\exp[2 \beta \Delta G_0]
\nonumber\\
\alpha_b &=& C^0_l \Psi^2 C^\circ K({\bf r})
\nonumber \\
\beta_b &=& -C^0_l C^\circ K({\bf r}) \Psi^2 \left( 
2 C^0_l + 2 C^\circ \exp[\beta \Delta G_0]
\right) \, \, ,
\label{eq:den2HCls}
\end{eqnarray}
where $N=4\pi R^2 \Psi$. We omit the lengthily expressions of $\gamma_1$, $\gamma_s$, and $\gamma_b$. Using Eqs.~\ref{eq:den1HCls}, \ref{eq:den2HCls} and Eq.~\ref{eq:free}, we find that $\Delta f_\mathrm{coll} = -n_b /2$ to  leading order. In particular, we find that 
\begin{eqnarray}
\beta \Delta f_\mathrm{coll} &=& - { Z C^0_l R \Psi^2 C^\circ v_0 \over 2 (C^0_l)^2 } \, \, .
\end{eqnarray}

\bibliography{2_SM.bbl}